\documentclass[letterpaper,11pt,onecolumn]{article}
\usepackage[margin=1in]{geometry}
\usepackage[hyphens]{url}
\usepackage{graphicx}
\usepackage{natbib}
\usepackage{caption}
\usepackage{booktabs}
\usepackage{microtype}
\usepackage{amsmath}
\usepackage{amssymb}
\usepackage{mathtools}
\usepackage{amsthm}

\usepackage{algorithm}
\usepackage{algpseudocode}

\theoremstyle{plain}

\theoremstyle{definition}

\theoremstyle{remark}

\newcommand{\affiliations}[1]{\date{#1}}

\title{Domain-Informed Multi-View Self-Distillation for Astronomical Light-Curve Representation Learning with JEPA}
\author{Yicheng Rui}
\affiliations{Tsung-Dao Lee Institute, Shanghai Jiao Tong University, Shanghai 201210, China\\ruiyicheng@sjtu.edu.cn}

\begin{document}

\maketitle

\begin{abstract}
Light curves describe temporal variations in the brightness of celestial objects. Learning robust representations of light curves is essential for large-scale automatic discovery in the dynamic universe, but existing time-series foundation models often struggle with the uneven sampling, complex noise, and wide range of physical timescales that characterize astronomical observations. We propose a domain-informed representation learning framework for irregular astronomical time series with Joint-Embedding predictive architecture (JEPA), combining semantics-preserving views, uncertainty-aware tokenization, and multi-view self-distillation. The model instantiates this framework with three views of each light curve: the raw sequence, the Generalized Lomb-Scargle periodogram, and the phase-folded light curve. Each view is processed by a separate encoder equipped with continuous Rotary Positional Embedding (C-RoPE) for irregular observation times and Error-Aware Numeric Embedding (EANE) for measurement uncertainties. The encoders are trained with multi-view self-distillation using LeJEPA regularization on the LEAVES dataset and evaluated on the StarEmbed classification benchmark. On StarEmbed, our model outperforms hand-crafted features on 15 of 16 classification metrics. In few-shot linear probing, it achieves macro-F1 scores of 42.56  $\pm$ 7.21 with one sample per class and 63.58 $\pm$ 1.20 with 100 samples per class, consistently improving over hand-crafted features. Beyond variable-star classification, the learned representation supports similarity search, parameter estimation, and photometric zero-point drift detection. We further evaluate cross-domain adaptation on 12 heterogeneous irregular time-series datasets from PYRREGULAR, where the adapted variant matches or exceeds previous state-of-the-art performance on 5 datasets, compared with at most 3 wins by any single prior baseline. These results demonstrate that domain-informed multi-view self-distillation is an effective strategy for learning representations of irregular time series, while also highlighting that successful time-series representation learning requires domain-specific inductive biases rather than a universally optimal architecture. Code of this project is available at \url{https://github.com/ruiyicheng/domain-informed-lightcurve-jepa} .
\end{abstract}

\section{Introduction}

Light curves describe temporal variations in the brightness of celestial objects. They are the primary data product for understanding the dynamic universe. With the advent of next-generation autonomous optical surveys, such as the Vera C. Rubin Observatory \citep{Ivezi19} and the Tianyu telescope \citep{Feng24}, we are entering an era of "Big Data" astronomy, expected to yield billions of celestial light curves annually \citep{Rui25}. Learning robust, global representations of these time series is a critical prerequisite for high-stakes downstream tasks like variable star classification \citep{Li25}.
Unlike other types of time series, the light curves have the following features:

\begin{enumerate}
 \item Extremely uneven sampling: For ground-based telescopes, light curves are extremely unevenly sampled because observations are limited to nighttime. This leads to inevitable gaps between nights and seasons. Meanwhile, the telescope schedule would also lead to extremely long gaps between observations. Some schedulers, e.g. Zwicky Transient Facility (ZTF), would also lead to uneven cadence (separation between observations) for intra-night observations \citep{Bellm19}. 
 \item Complicated noise: Noise in light curve comes from the quantum nature of light, the atmospheric turbulence (scintillation), changes in weather conditions, and the instruments \citep{Rui_2026_CVPR}. Empirically, noise of the light curve is temporally correlated \citep{li2026astroskyflowastronomicalskyimage}. For faint sources, the noise of a data point in the light curve can be larger than the signal itself, e.g., for a transiting exoplanet \citep{Zheng_2026}. Phase-folding and binning are required to obtain a significant signal.
 \item Large dynamic range of timescales: Astronomical phenomena span different timescales. For example, the period of eclipsing binaries can range from several minutes to tens of years. What's worse, some types of variable stars have multiple variation periods in the light curve across different time scale. Therefore, it is necessary to capture the signal in the light curve with different timescale for the downstream tasks.
\end{enumerate}

Due to these features, "off-the-shelf" Transformer-based models often underperform for the light-curve representation learning. Specifically, vanilla time embeddings fail to capture the underlying sampling density, and auto-regressive pre-training (GPT-style) tends to overfit when faced with high-noise sequences \citep{Wang24} like light curve. Therefore, most light curve representation models, including  general-purpose time series foundation models (TSFMs) like Moirai \citep{liu2026moirai20timeseries}, obtain worse performance than hand-crafted features in light curve embedding learning benchmarks such as StarEmbed  benchmark \citep{Li25}. 

 

To address these problems, we propose a novel approach for learning the global representation of astronomical light curves with Joint-Embedding Predictive Architecture (JEPA) \citep{lecun2022}, which applies regularization across multiple views of the light curve for self-distillation. For representing the sampling times of light curves with the prior of time-translation invariance for light curves, we introduce the continuous adaptation of Rotary
Positional Embedding (C-RoPE). For representing the uncertainty of light-curve measurements, we introduce the Error-Aware Numeric Embedding (EANE). For training the representation model, we employ a multi-view self-distillation technique with the LeJEPA regularization loss \citep{Balestriero25}. 
The model achieves advantage over hand-crafted features on 15 out of 16 metrics on StarEmbed classification benchmark. To the best of our knowledge, this is the first foundation model that defeats the hand-crafted features in this classification benchmark. Besides, our model can also be used for applications including similarity search, photometric zero-point drift detection and the parameter estimation of light curve. Furthermore, our adapted model achieved SOTA performance on 5 of the 12 unevenly sampled datasets in PYRREGULAR after fine-tuning, indicating the potential of our model architecture for learning the representation of time series beyond astronomy.

This paper is organized as follows. In Section \ref{sec:method}, we introduce the methods used in our model; in Section \ref{sec:dab}, we introduce the dataset and benchmark used in this work; in Section \ref{sec:exp}, we introduce the experimental results of our model; in Section \ref{sec:app}, we introduce the applications including similarity search, parameter estimation, and the applications in other domains; in Section \ref{sec:rw}, we give a literature review of related work to illustrate our contribution; in Section \ref{sec:cad}, we summarize the conclusion of the paper and discusses the results.

\section{Method \label{sec:method}}
\subsection{Architecture of the model}

\begin{figure*}
 \centering
 \includegraphics[width=\linewidth]{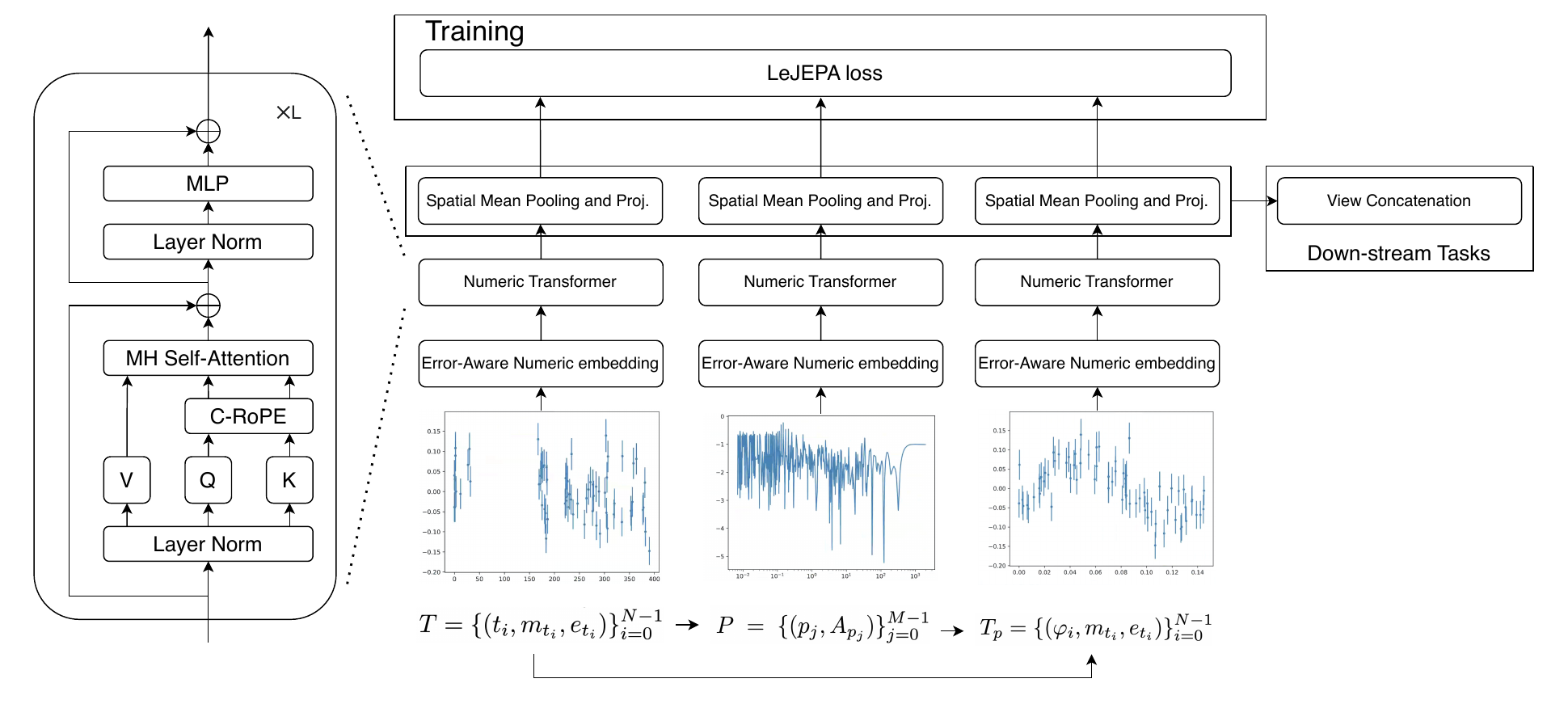}
 \caption{Architecture of proposed model in this work.  Arrows refer to the direction of forward process.}
 \label{fig:architecture}
\end{figure*}
The model architecture used in this work is shown in Fig. \ref{fig:architecture}. 
The input of the model is the raw light curve with $N$ measurements represented as $T = \{(t_i,m_{t_i},e_{t_i})\}_{i=0}^{N-1}$, where $t_i$ is the time of the $i$th sample, $m_{t_i}$ is the measurement of the $i$th data point and $e_{t_i}$ is the uncertainty. The raw light curve is transformed into domain-inspired views as described in Section \ref{sec:views}. For each view, a transformer numeric encoder equipped with C-RoPE (see Section \ref{sec:crope}) and EANE (see Section \ref{sec:eane}) is used for encoding the view data. For inference, the embeddings from multiple views are concatenated for downstream tasks. For training, the LeJEPA loss is applied on a batch of light curves \citep{Balestriero25}. The LeJEPA loss is written as
\begin{equation}
 \begin{aligned}
\mathcal{L}_{\rm LeJEPA} &= \lambda \mathcal{L}_{\rm SIGReg} + (1 - \lambda) \mathcal{L}_{\rm inv} \\
 \mathcal{L}_{\rm inv} &= \frac{1}{V} \sum_{v=1}^{V} \| z_v - \bar{z} \|^2 \\
 \mathcal{L}_{\rm SIGReg} &= \mathbb{E}_{\bar{z}} \left[ \int \left| \hat{\phi}_{\rm data}(t; \bar{z}) - \phi_{\rm target}(t) \right|^2 w(t) \, dt \right],
 \end{aligned}
\end{equation}
where $V=3$ denotes the number of views; $z_v$ represents the latent embedding obtained by the encoder for view $v$; and $\bar{z} = \frac{1}{V} \sum_{v=1}^{V} z_v$ is the centroid of the embeddings across all views. The terms $\hat{\phi}_{\rm data}$ and $\phi_{\rm target}$ refer to the empirical characteristic function of the projected features and the characteristic function of the target standard Gaussian distribution, respectively; $w(t)$ is a weight function with standard Gaussian distribution.
The LeJEPA loss function consists of two parts. $\mathcal{L}_{\rm inv}$ is the multi-view alignment loss function that makes the alignment of semantics between all domain-informed views of a light curve; $\mathcal{L}_{\rm SIGReg}$ is the regularization loss to make the latent of light curves distributed as a  standard Gaussian distribution, which is optimal for linear probing and radial-based k-NN \citep{Balestriero25}. This architecture enables the unsupervised learning of light curve representation. Appendix \ref{app:alg} provides a summary of the algorithm.

\subsection{Views of light curves\label{sec:views}}
In this work, we consider three views of the light curves

\begin{enumerate}
 \item The raw data of the light curve $T$;
 \item The generalized Lomb-Scargle periodogram (GLS) power spectrum $P = \{(p_j,A_{p_j})\}_{j=0}^{M-1}$, where $p_j$ is the $j$th period, and $A_{pj}$ is the corresponding power.  In this work, GLS \citep{Zechmeister09} is used for extracting the power spectrum of the light curve. Details of the implementation of GLS is shown in Appendix \ref{sec:GLS}.
 \item The phase-folded light curve  \\$T_p = \{(\varphi_i,m_{t_i},e_{t_i})\}_{i=0}^{N-1}$, where 
\begin{equation}
   \varphi_i = \frac{t_i}{P_o} -  \left\lfloor \frac{t_i}{P_o} \right\rfloor ; \quad  P_{\rm o} = {\rm argmax}_P {\rm GLS}(P),
\end{equation}
which would help extracting the morphology of periodic light variation.
\end{enumerate}
The choice of these views is from the domain-specific knowledge of light curve diagnosis. The raw light-curve view is mainly used for global trend diagnosis, variable amplitude determination, and long-period quasi-periodic light curve diagnosis; the GLS power spectrum is used for period extraction and diagnosis for targets with multiple oscillation modes; the phase-folded light curve is used for extracting the morphology of the oscillation, because the period of the target can be even shorter than the observational cadence. We assume these three views have similar semantics because all of them are derived from the raw light curve data. 

\subsection{Continuous Rotary Positional Embeddings (C-RoPE)\label{sec:crope}}

Unlike traditional sequence models that utilize discrete position indices, we employ a continuous adaptation of Rotary Positional Embeddings (C-RoPE) to account for the arbitrary temporal spacing of astronomical observations. Given a continuous coordinate $t \in \mathbb{R}$ (e.g., time or frequency), we map it to a complex rotation in the hidden space.

We define a set of angular frequencies $\omega_d$ for each dimension pair $d \in \{1, \dots, D/2\}$ using a base frequency $\omega_0$:
\begin{equation}
 \omega_d = \omega_0^{\frac{-2(d-1)}{D}}
\end{equation}

For a specific coordinate $t$, the phase $\theta_{t,d} = t \cdot \omega_d$ is used to construct the embedding vectors $\mathbf{c}_t = \cos(\mathbf{\Theta}_t)$ and $\mathbf{s}_t = \sin(\mathbf{\Theta}_t)$, where $\mathbf{\Theta}_t$ is the concatenated phase vector. The rotary transformation is applied to the query ($\mathbf{q}$) and key ($\mathbf{k}$) vectors as follows:
\begin{equation}
 f(\mathbf{x}, t) = \mathbf{x} \odot \mathbf{c}_t + \text{rotate\_half}(\mathbf{x}) \odot \mathbf{s}_t
\end{equation}
where $\odot$ denotes the Hadamard product and 
\begin{equation}
\text{rotate\_half}([x_1, \dots, x_{D/2}, x_{D/2+1}, \dots, x_D]) = [-x_{D/2+1}, \dots, -x_D, x_1, \dots, x_{D/2}]
\end{equation}
. This formulation ensures that the attention mechanism is invariant to absolute shifts in $t$ and only depends on the relative interval $\Delta t = t_i - t_j$. This encodes astronomical prior knowledge.

\subsection{Error-Aware Numeric Embedding (EANE) \label{sec:eane}}

To account for measurement uncertainty, we introduce an Error-Aware Numeric Embedding (EANE) layer. Rather than treating a magnitude $m_{t_i}$ as a point estimate, we treat it as a probability distribution $p(m_{t_i}) = \mathcal{N}(m_{t_i}, e_{t_i}^2)$.

The continuous value $m_{t_i}$ is first mapped to a discrete bin index via a quantization function $Q(m_{t_i}) \in \{0, \dots, N_{\rm bin}-1\}$ over the range $[m_{min}, m_{max}]$. To propagate the uncertainty $e_{t_i}$ through the embedding layer ${\rm Emb}$, we compute the expected embedding vector $\mathbf{z}_{t_i}$ by integrating over the Gaussian PDF:
\begin{equation}
 \begin{aligned}
 \mathbf{z}_{t_i} &= \mathbb{E}_{\epsilon \sim \mathcal{N}(0,1)} [{\rm Emb}(Q(m_{t_i} + e_{t_i} \epsilon))]\\
 & = \int_{-\infty}^{\infty} \frac{1}{\sqrt{2\pi}} e^{-\frac{\epsilon^2}{2}} {\rm Emb}(Q(m_{t_i} + e_{t_i} \epsilon)) \mathrm{d}\epsilon\\
 & \approx \frac{\sum_{k=-R}^{R} \exp\left( -\frac{1}{2} \left( \frac{c_{Q(m_{t_i}) + k} - m_{t_i}}{e_{t_i}} \right)^2 \right) {\rm Emb}(Q(m_{t_i}) + k)}{\sum_{k=-R}^{R} \exp\left( -\frac{1}{2} \left( \frac{c_{Q(m_{t_i}) + k} - m_{t_i}}{e_{t_i}} \right)^2 \right) },
 \end{aligned}
\end{equation}
where  $c_\cdot$ is the bin center of $Q^{-1}(\cdot)$.  For simplicity, the uncertainty is not considered in the GLS view, i.e. $e_{t_i}\to 0$ in EANE.

\section{Dataset and benchmark\label{sec:dab}}

In this work, we adopt LEAVES as the source for pretraining \citep{Fei_2024}. LEAVES is a compatible, multi-survey light-curve dataset curated from publicly available time-series photometry, integrating data from the ASAS-SN Catalog of Variable Stars, Gaia DR3, and ZTF DR2. 
 The current LEAVES DR1 release contains $\sim 980$k mono-band variable-star light curves (organized into 6 superclasses and 9 subclasses) and $\sim130$k non-variable light curves, supporting both 7-class  and 10-class (subclasses + non-variable) classification setups. The LEAVES dataset is an ideal source for light curve pretraining.

StarEmbed \citep{Li25} is a benchmark for TSFMs on variable star light curves. It contains $\sim$40k confirmed variable stars from  Catalina Surveys
Periodic Variable Star Catalog (CSPVS) from ZTF multi-band light curves \citep{Drake14}. The downstream tasks include clustering, classification, and out-of-distribution detection, which is an ideal playground for testing light curve embedding models. Interestingly, the hand-crafted features are highly competitive in all tasks of StarEmbed benchmark. Especially, the hand-crafted features reach almost all the state-of-the-art performance in the classification metrics. 
Before pretraining, we cross-matched LEAVES and StarEmbed using Gaia DR3 source identifiers and removed all LEAVES objects that matched any StarEmbed source.

To test the capability of model beyond astronomical light curves, we test our model on PYRREGULAR benchmark \citep{spinnato2026pyrregularunifiedframeworkirregular}. It provides classification benchmarks for 12 unevenly sampled datasets with 12 baselines. In this work, we retrain our model  on the training set, and use Linear and k-NN heads to evaluate classification performance.

\section{Experiments \label{sec:exp}}
\subsection{Baseline and ablation}
To compare the performance of our model against previously established models, and study the effects of different components in the model, we set up a series of baseline and ablation experiments. The baseline contains domain-specific light curve feature extractors and general-purpose time series foundation models.   For domain-specific light curve feature extractors  like Astromer-2 \citep{Donoso25} and FALCO \citep{Zuo26}, we pretrained the following models on the LEAVES dataset to exclude the  dataset effects. Details of the training and model parameter of these baselines are shown in Appendix \ref{app:model_parameters_experimental_details}.
\begin{itemize}
 \item Astromer-2 (BERT-like): As a domain-specific model for astronomical light curve, the Astromer-2 model uses the continuous-time position embedding as the positional embedding. Astromer-2 uses a mask-reconstruction loss for light curve modeling. Meanwhile, it utilizes the uncertainty of light curve in the reconstruction loss.  In the original paper, the data of MACHO survey \citep{Alcock01} is used for training. It achieved poor performance in the StarEmbed benchmark because it has limited ability to transfer between original dataset and the ZTF light curve used by the original paper. For fair comparison, we re-pretrain the model on LEAVES dataset.
 \item FALCO (GPT-like): FALCO uses a GPT-like architecture and auto-regressive loss for pre-training. Because the model is not available to public, we pretrained our own version of FALCO using the LEAVES data.
 \item Chronos (General-purpose Time-series SOTA) \citep{Fatir24}: In this work, we use the latest version of Chronos model, Chronos-2, together with Chronos-tiny-bolt, which is used in StarEmbed benchmark, as the baseline for general-purpose time series foundation model. 
 \item Hand-crafted features: Hand-crafted features are a highly competitive feature for variable star-related tasks. This is partially because the original taxonomy of variable star is based on the morphology of light curve, therefore using a decision-tree like method for classification by definition. We reproduce the extraction of hand-crafted features of StarEmbed benchmark, which is based on FATS \citep{Nun15}. 
\end{itemize}

To illustrate the effects of C-RoPE, error-aware numeric embedding, and  multi-view self-distillation, we also setup several ablation experiments to illustrate the effects of the components. They are 

\begin{itemize}
 \item Positional Embedding ablation: To study the effects of C-RoPE, we implement the model with the same setup but only using the classical rotational positional embedding \citep{2021arXiv210409864S} where the phase $\theta_{i,d} = i\cdot\omega_d$ only depends on the position in the light curve. Therefore, the observational time $t_i$ is not used in this ablation.
 \item Embedding uncertainty ablation: To study the effects of error-aware numeric embedding, we setup a model that does not consider the uncertainty of the input light curve, i.e. all probability mass is concentrated in one quantized bin that is located in the value of light curve. 
 \item Loss ablation: To study the effects of the LeJEPA loss, we setup models that  use symmetric CLIP loss between raw light-curve view and other two views for comparison.
 \item View ablation: To study the effects of different views, we setup ablation experiments  that exclude raw light-curve view, the GLS periodogram view, and the phase-folding view. 
\end{itemize}
To test the capability of our model in other domains, we also test our model on all 12 irregularly sampled datasets within PYRREGULAR  \citep{spinnato2026pyrregularunifiedframeworkirregular}.

In this work, all training, evaluation, and testing were conducted on a machine equipped with eight 32 GB NVIDIA V100 GPUs, two Intel Xeon E5-2690 processors, and 128 GB of system memory. 



\subsection{StarEmbed classification benchmark}

 We compare the performance of our model with the baselines. The results are shown in Table \ref{tab:classificationres}. Our full model achieved superior performance  on 15 out of 16 classification metrics in StarEmbed benchmark relative to the hand-crafted feature. 
 
  \begin{table*}[t]
    \centering
    \caption{Classification results of the StarEmbed benchmark in percentile. For fair comparison, Astromer-2 and FALCO
    are re-pretrained using LEAVES dataset. Without PF refers to model with no phase-folded view. The upper panel shows
    the baseline model results and the lower panel shows the ablation experiments results. The best-performing model in each column is in bold and the second best is underlined. Values are rounded to one decimal place; win counts are computed from unrounded scores.}
    \label{tab:classificationres}
    \scriptsize
\setlength{\tabcolsep}{3pt}
\renewcommand{\arraystretch}{0.95}
    \begin{tabular}{l|cccc|cccc|cccc|cccc}
    \specialrule{2pt}{1pt}{1pt}
    Model
    & \multicolumn{4}{c|}{MLP}
    & \multicolumn{4}{c|}{$k$-NN}
    & \multicolumn{4}{c|}{Logistic}
    & \multicolumn{4}{c}{Random Forest} \\
    \cline{2-17}
    & Acc & Rec & Prec & F1
    & Acc & Rec & Prec & F1
    & Acc & Rec & Prec & F1
    & Acc & Rec & Prec & F1 \\
    \specialrule{2pt}{1pt}{1pt}
    Hand-crafted
    & 81.0 & 83.7 & 67.0 & 71.9
    & 89.1 & 68.2 & \underline{81.1} & 72.3
    & 80.2 & 82.4 & 63.6 & 67.7
    & \underline{90.4} & 69.8 & \underline{86.4} & 74.4 \\
    \hline
    Chronos-2
    & 77.1 & 76.5 & 57.0 & 63.1
    & 84.9 & 59.7 & 72.9 & 62.8
    & 77.8 & 70.3 & 58.8 & 62.7
    & 86.6 & 58.2 & 75.3 & 61.6 \\
    \hline
    Chronos-Bolt
    & 73.3 & 72.2 & 55.6 & 60.5
    & 80.7 & 54.2 & 64.8 & 56.9
    & 70.8 & 67.5 & 54.7 & 57.8
    & 82.4 & 54.4 & 70.2 & 57.9 \\
    \hline
    Astromer-2$^*$
    & 75.4 & 62.3 & 48.9 & 52.4
    & 81.5 & 44.0 & 51.6 & 45.2
    & 69.1 & 61.3 & 47.0 & 49.4
    & 83.6 & 46.2 & 52.8 & 47.6 \\
    \hline
    FALCO$^\dagger$
    & 77.3 & 71.8 & 54.0 & 59.8
    & 81.3 & 48.9 & 69.6 & 52.8
    & 75.4 & 73.9 & 56.3 & 60.8
    & 84.1 & 51.6 & 64.9 & 55.5 \\
    \specialrule{2pt}{1pt}{1pt}
    Without C-RoPE
    & 84.8 & 83.1 & 65.6 & 71.2
    & 87.0 & 67.5 & 76.9 & 70.8
    & 85.3 & 83.7 & 67.9 & 73.1
    & 89.0 & 68.2 & 84.3 & 72.7 \\
    \hline
    Without EANE
    & 84.4 & 85.0 & 64.8 & 71.6
    & 89.8 & 71.8 & 76.8 & 73.4
    & 85.1 & 80.6 & 66.3 & 71.2
    & 90.1 & 69.2 & 80.4 & 72.2 \\
    \hline
    Without raw
    & \underline{85.7} & 86.0 & 65.7 & 72.3
    & 89.3 & \underline{75.0} & 79.9 & \underline{77.1}
    & 85.2 & \underline{84.2} & 68.3 & 73.4
    & 90.3 & \textbf{73.8} & 84.8 & \textbf{77.8} \\
    \hline
    Without PF
    & 84.8 & 83.4 & 64.8 & 71.5
    & 87.9 & 70.0 & 77.6 & 72.7
    & 85.1 & 84.1 & 65.8 & 72.0
    & 88.9 & 67.1 & 85.0 & 71.1 \\
    \hline
    Without GLS
    & 80.1 & 79.2 & 60.5 & 65.5
    & 87.9 & 65.7 & 76.0 & 68.7
    & 78.1 & 78.0 & 59.5 & 63.9
    & 88.8 & 64.9 & 80.5 & 68.9 \\
    \hline
    CLIP loss
    & 85.4 & \textbf{86.9} & \underline{67.9} & \textbf{74.2}
    & \underline{90.1} & 73.6 & \textbf{82.9} & 76.6
    & \textbf{86.4} & \textbf{84.8} & \underline{68.4} & \underline{74.0}
    & 90.1 & 68.2 & \textbf{88.1} & 73.7 \\
    \specialrule{2pt}{1pt}{1pt}
    Our full model
    & \textbf{86.5} & \underline{86.6} & \textbf{68.0} & \underline{74.1}
    & \textbf{90.2} & \textbf{75.2} & \underline{81.1} & \textbf{77.6}
    & \underline{86.3} & 84.0 & \textbf{69.7} & \textbf{74.4}
    & \textbf{90.5} & \underline{73.4} & 85.0 & \underline{77.4} \\
    \specialrule{2pt}{1pt}{1pt}
    \end{tabular}
  \end{table*}

 To test the few-shot classification capability of our model, we run the linear-probe classification using 1/5/20/100 samples per class from the training set, and test the results in the full test set. The few-shot samples from the StarEmbed training set are bootstrapped for 2000/400/100/20 times respectively for the uncertainty of the few-shot classification metrics. The F1 results are shown in Table \ref{tab:fewshot_bootstrap_classification_f1}. It is shown that our model achieves consistent improvement over hand-crafted features. in few-shot classification experiments.
 Details of these experiments are shown in Appendix \ref{app:model_parameters_experimental_details}.

\begin{table*}[t]
   \caption{Few-shot bootstrap linear-classification macro F1 (in percentile, mean $\pm$ uncertainty) across bootstrap.}
   \label{tab:fewshot_bootstrap_classification_f1}
   \centering
   \begin{tabular}{lcccc}
   \specialrule{2pt}{1pt}{1pt}
   Model & $n=1$ & $n=5$ & $n=20$ & $n=100$ \\
   \specialrule{2pt}{1pt}{1pt}
   Hand-crafted & 30.28 $\pm$ 5.94 & 47.73 $\pm$ 4.91 & 56.43 $\pm$ 2.96 & 62.39 $\pm$ 1.70 \\
   Chronos-2 & 31.77 $\pm$ 6.41 & 45.35 $\pm$ 3.79 & 50.86 $\pm$ 2.34 & 53.55 $\pm$ 0.92 \\
   Chronos-Bolt & 28.63 $\pm$ 5.50 & 39.57 $\pm$ 3.48 & 45.11 $\pm$ 2.32 & 49.08 $\pm$ 1.03 \\
   Astromer-2 & 18.15 $\pm$ 4.74 & 30.39 $\pm$ 3.23 & 37.35 $\pm$ 1.83 & 41.84 $\pm$ 0.79 \\
   FALCO & 21.74 $\pm$ 4.74 & 33.12 $\pm$ 3.31 & 42.30 $\pm$ 2.68 & 52.13 $\pm$ 1.33 \\
   \specialrule{2pt}{1pt}{1pt}
   Without C-RoPE & 29.98 $\pm$ 5.98 & 43.40 $\pm$ 3.97 & 52.40 $\pm$ 2.94 & 59.54 $\pm$ 0.85 \\
   Without EANE & \underline{39.84 $\pm$ 7.17} & \underline{53.94 $\pm$ 4.51} & 58.45 $\pm$ 2.44 & 60.43 $\pm$ 1.23 \\
   Without raw & 37.78 $\pm$ 6.92 & 52.15 $\pm$ 4.36 & 59.00 $\pm$ 2.82 & 62.69 $\pm$ 1.36 \\
   Without PF & 29.47 $\pm$ 6.21 & 44.79 $\pm$ 4.04 & 53.72 $\pm$ 2.23 & 59.88 $\pm$ 1.25 \\
   Without GLS & 27.04 $\pm$ 5.51 & 41.25 $\pm$ 5.29 & 47.69 $\pm$ 3.54 & 52.75 $\pm$ 1.43 \\
   CLIP loss & 33.19 $\pm$ 6.50 & 50.75 $\pm$ 4.23 & \underline{59.45 $\pm$ 2.54} & \underline{62.88 $\pm$ 1.43} \\
   \specialrule{2pt}{1pt}{1pt}
   Our full model & \textbf{42.56 $\pm$ 7.21} & \textbf{54.56 $\pm$ 4.07} & \textbf{60.04 $\pm$ 2.63} & \textbf{63.58 $
   \pm$ 1.20} \\
   \specialrule{2pt}{1pt}{1pt}
   \end{tabular}
  \end{table*}

\subsection{Qualitative Interpretability}

\begin{figure}[h]
 \centering
 \includegraphics[width=0.8\linewidth]{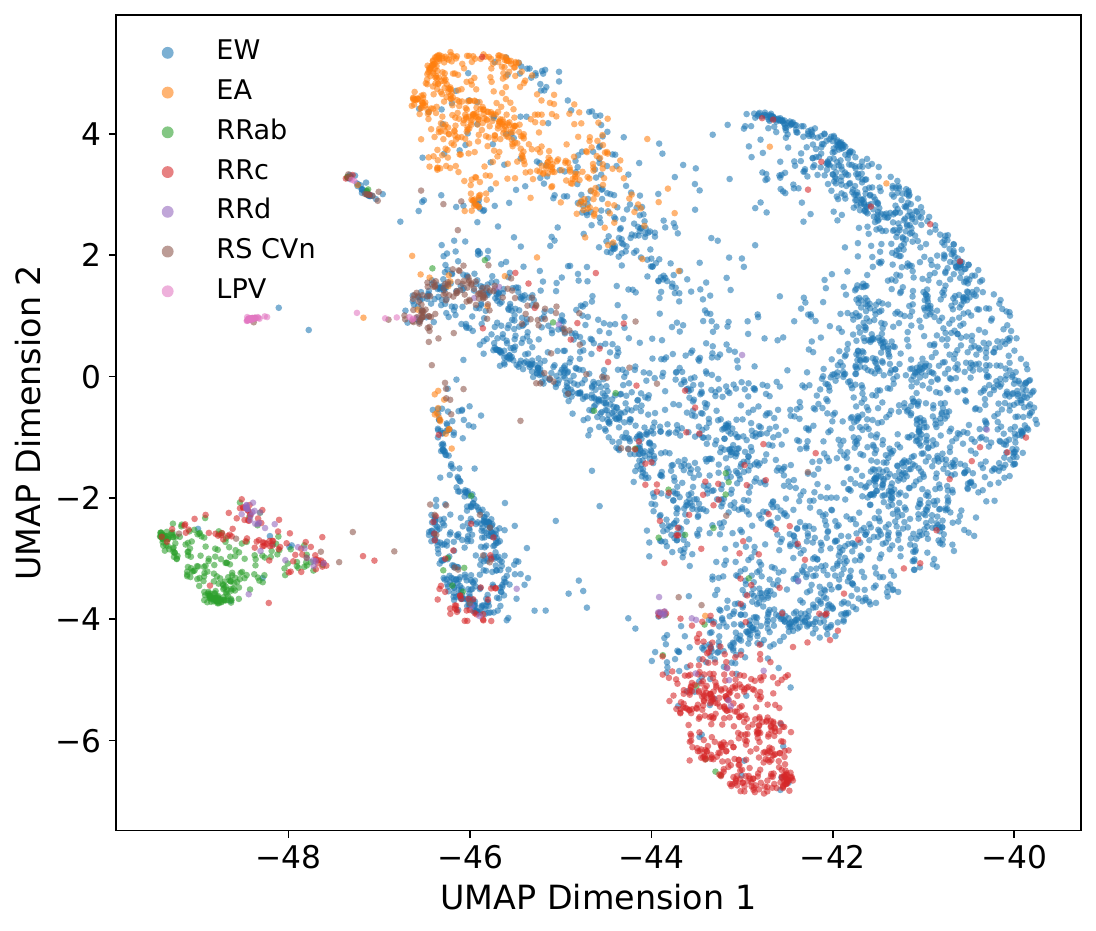}
 \caption{Two-dimensional UMAP projection of the StarEmbed test set embedding. One dot represents a sample in the StarEmbed test set.}
 \label{fig:umap_Test_Split}
\end{figure}

The two-dimensional Uniform Manifold Approximation and Projection (UMAP)  projection \citep{McInnes18}  of StarEmbed test set embedding of our model is shown in Fig. \ref{fig:umap_Test_Split}. It is shown that variable stars of the same class are clustered together in the UMAP projection, which indicates that our unsupervised embedding model would lead to well-behaved geometric distribution for resulting embeddings. 

\begin{figure}[h]
 \centering
 \includegraphics[width=\linewidth]{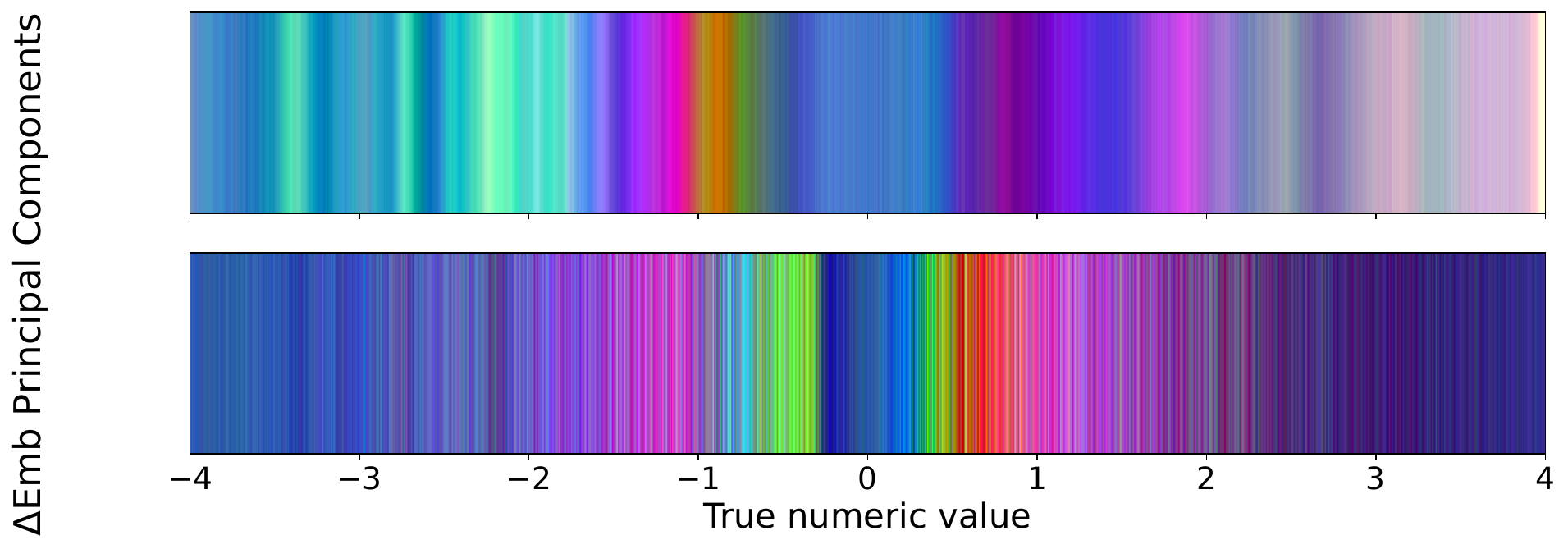}
 \caption{Visual comparison of the first three principal components of $\Delta$Emb  from the full model (upper) and the ablation model with no EANE (lower). The first three components are represented by red, green and blue. The synthesized color represents the proportion of the three principal components.}
 \label{fig:PCAemb}
\end{figure}

To clarify the effects of EANE, we compare the embedding for the raw light-curve view of our model and the ablation model that concentrates all the probability mass to the bin of the corresponding numeric value. We visualize the first three principal components of the change of embedding during the training $\Delta {\rm Emb} = {\rm Emb_{\rm final} - Emb_{\rm init}}$ for different input values of the raw light-curve view embedding in Fig. \ref{fig:PCAemb}. It is shown that the model that use EANE have a smoother embedding training progress, which is a desirable feature of EANE because nearby numeric value of the light curve points have similar semantic meaning. 

\section{Application\label{sec:app}}
\subsection{Similarity search}



\begin{figure*}
 \centering
 \includegraphics[width=\linewidth]{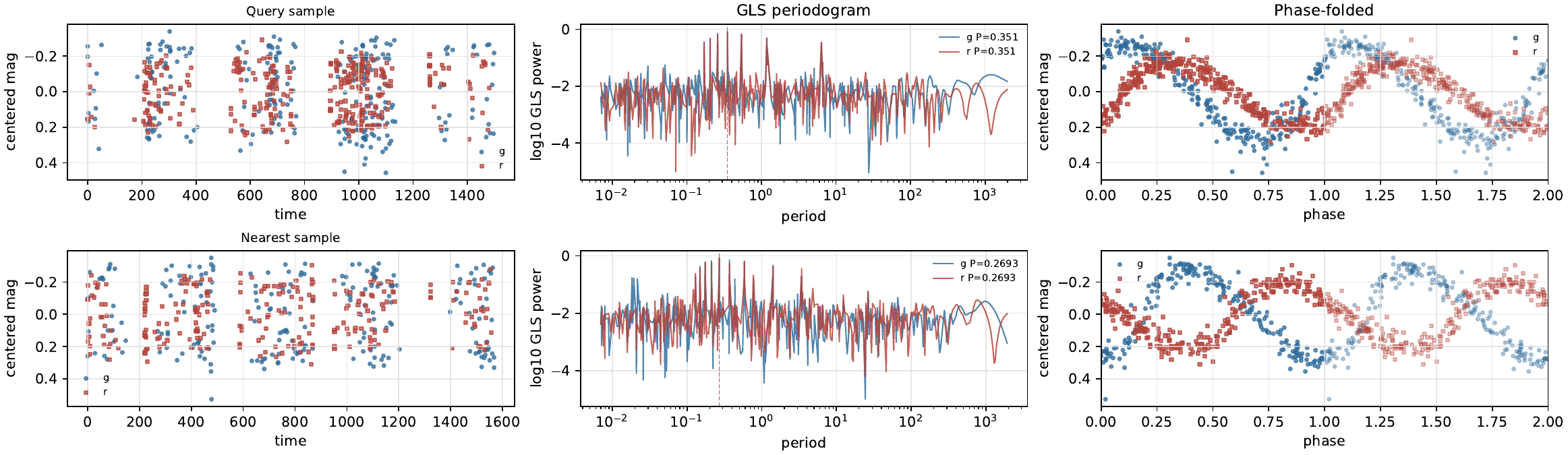}
 \caption{An example for similarity search. The upper panel shows the three views of a target light curve; the lower panel shows the light curve with highest similarity (lowest L2 distance) in the StarEmbed test set.}
 \label{fig:rrccossim}
\end{figure*}

Our pretrained model exhibits strong capability for similarity search. For an encoded light curve represented by the embedding vector $\vec{z}$, we can retrieve the most similar light curves based on the L2 distance $D$ between samples, defined as
\begin{equation}
 D_{ij} = \lVert \vec{z}_i - \vec{z}_j \rVert.
\end{equation}
For a pair of similar light curves $i$ and $j$, the L2 distance should be small. An example of similarity search is shown in Fig. \ref{fig:rrccossim}. This figure shows an example of similarity search. An RRc-type variable star is used as the search target. It is shown that the search result succeeded in retrieving a similar light curve that belongs to the same class with similar morphological features. In astronomical research, this can help astronomers with population studies and dataset collection. 

\subsection{Parameter estimation}
\begin{figure}
 \centering
 \includegraphics[width=1\linewidth]{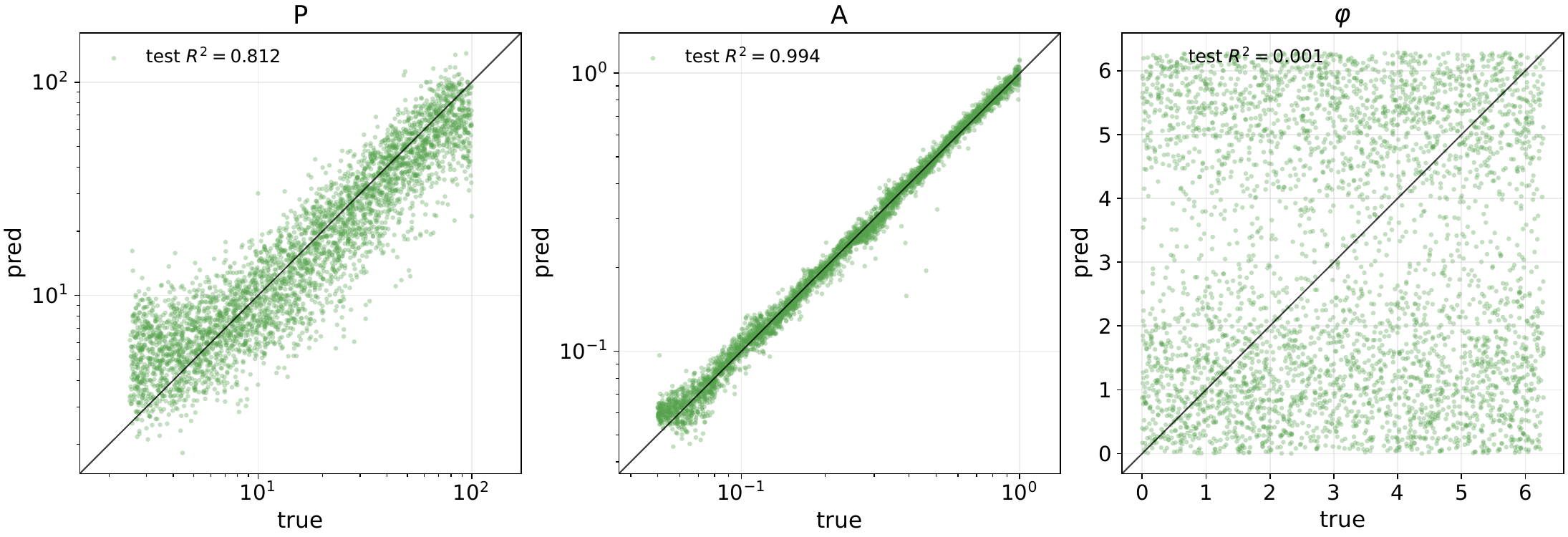}
 \caption{The recovery of sinusoidal signals on the test set. Each dot represents a sample from the test set. }
 \label{fig:true_vs_pred_recovery}
\end{figure}

With linear probe fine-tuning, our model can be used as an initial guess to accelerate Bayesian parameter estimation. An injection-recovery test is conducted to illustrate the parameter estimation capability of our model. We generated 20,000 light curves with a sinusoidal signal with white noise
\begin{equation}
 m_{t_i} = A \sin (\frac{2\pi}{P} t_i + \varphi) + e\epsilon,
\end{equation}
where period $P$ is sampled log-uniformly from 2.5 days to 100 days; amplitude $A$ is sampled log-uniformly from 0.05 mag to 1 mag; phase $\varphi$ is sampled uniformly from 0 to $2\pi$; observation time $t_i$ is sampled from the LEAVES dataset with more than 100 observations; $\epsilon$ is sampled from standard normal distribution $\mathcal{N}(0,1)$; and  the uncertainty of light curve $e$ is set to be 0.01 magnitude. The 20,000 light curves are split to train, validate, and test set in 7:1:2. The training has 500 epochs. We choose the epoch with lowest validation loss. A linear probe on the embedding is trained to fit $P$, $A$, and $\varphi$. The recovery result is shown in Fig. \ref{fig:true_vs_pred_recovery}. It is shown that the period $P$ and amplitude $A$ are recovered  while the linear probe failed to recover the phase $\varphi$.  This is because in our model, all the time embeddings use relative time. This makes our model time-translation invariant. The phase information is therefore not encoded in the embedding. This result aligns with the philosophy of Joint-Embedding Predictive Architecture (JEPA) that the embedding should be a summary of input data instead of the encoded data that contains all the details of input data \citep{lecun2022}.

\subsection{Photometric zero-point drift detection}
The photometric system may undergo a drift of photometry zero point due to instrument changes, telescope pointing issues, or updates to the calibration pipeline. Our pretrained model can be used to detect this drift. To illustrate the model's capability for detecting this kind of anomaly, we injected the following signal $\Delta m$ into each light curve in the StarEmbed dataset
\begin{equation}
 \Delta m =\eta  d H(t_i - t_{\rm drift}),
\end{equation}
where $H(\cdot)$ is the Heaviside step function, 
\begin{equation}
 p(\eta) = \begin{cases}
\frac{1}{2}, & \eta = 1,\\
\frac{1}{2}, & \eta = -1,\\
0, & \text{otherwise},
\end{cases}\quad d\sim \mathcal{U}(d_{\rm min},d_{\rm max}),
\end{equation}
$t_{\rm drift}$ is sampled from 20th to 80th percentile of the observation time. 
The dataset consists of the original light curves with injected light curves. A logistic regression is applied on the features from our model to distinguish whether a zero-point drift has been injected into a light curve. The classification F1 for different zero-point drift $(d_{\rm min},d_{\rm max})$ and different view ablation is shown in Table \ref{tab:sudden_jump_model_sweep_test_f1}. It is shown that all views and the EANE make contributions to the photometric zero-point drift detection in the light curve, especially when the drift is small.



  \begin{table}[t]
      \caption{Test F1 for photometric zero-point drift detection.}
      \label{tab:sudden_jump_model_sweep_test_f1}
      \centering
      \begin{tabular}{l|cccc}
      \hline
      $d_{\rm min},d_{\rm max}$ & $0.05, 0.1$ & $0.1, 0.3$ & $0.3, 0.5$ & $0.5, 1.0$ \\
      \hline
      Without EANE & 0.7845 & 0.9430 & 0.9897 & 0.9945 \\
      Without Raw & 0.8049 & 0.9545 & 0.9920 & 0.9959 \\
      Without GLS & 0.7745 & 0.9090 & 0.9720 & 0.9896 \\
      Without PF & 0.7925 & 0.9509 & 0.9917 & 0.9963 \\
      Full model & \textbf{0.8292} & \textbf{0.9660} & \textbf{0.9936} & \textbf{0.9964} \\
      \hline
      \end{tabular}
  \end{table}

\subsection{Application in non-astronomical domains}

To explore the capability of the proposed framework beyond astronomy, we evaluate our model on the 12 irregularly sampled datasets in PYRREGULAR. Since time series from other domains have different value ranges, channel semantics, and sampling mechanisms from astronomical light curves, we apply the adaptation techniques described in Appendix~\ref{app:pyrregularadaption}. We fine-tune the model on the training split of each dataset for 10 epochs, and evaluate the learned representations using linear and k-NN classifiers. The full F1 comparison with previous state-of-the-art methods is reported in Table \ref{tab:PYRREGULAR_domain_adaptation} of Appendix~\ref{app:pyrregularadaption}.

The astronomy-informed model achieves state-of-the-art results on 3 of the 12 PYRREGULAR datasets. With the PYRREGULAR-adapted variant, the model matches or exceeds previous state-of-the-art performance on 5 datasets, while the best individual prior baseline in PYRREGULAR achieves state-of-the-art results on only 3 datasets. This result suggests that the proposed multi-view self-distillation framework can be competitive beyond astronomical light curves after domain adaptation. Meanwhile, the non-uniform gains across datasets also support our argument that irregular time-series representation learning requires domain-specific view and encoder design, rather than a universally optimal architecture.

\section{Related work\label{sec:rw}}
Most current time-series models, e.g. Chronos \citep{Fatir24,ansari2025chronos2univariateuniversalforecasting}, TimesFM \citep{2023arXiv231010688D}, TimeGPT \citep{garza2024timegpt1}, and Moirai \citep{liu2026moirai20timeseries}, focus on time series forecasting. One can use the mean-pooling or the CLS token of the last activation layer for the global representation \citep{elghoussani2026timeseriesfoundationmodelembeddings}. These models are trained on dataset from multiple domains with synthesized noise data. Chronos-2 supports forecast uncertainty as output. 
For work focusing on time series representation learning, models like TimesCLIP use contrastive CLIP alignment between natural language and the raw data \citep{2025arXiv250624124D}; TS-JEPA use an EMA teacher-student self distillation for learning the representation \citep{2025arXiv250925449E}.  However, most current TSFMs does not support input uncertainty, which is an important information loss in light curve analysis. 

Most previous works on light curve modeling use class labels as supervision signals, which is limited by the sample size and class impurity \citep{Naul_2017,Zuo26}. There are several works like FALCO \citep{Zuo26} and Astromer \citep{Donoso25} trying to learn the representation of light curve in an auto-regressive manner.

All the mentioned methods are limited to the raw light-curve view prediction or reconstruction, which cannot effectively extract phase-folded morphology.

\section{Conclusion and Discussion \label{sec:cad}}

In this work, we propose domain-informed multi-view self-distillation for astronomical light-curve representation learning with JEPA. For modeling the light curves with extremely unevenly sampled, noisy, and large range of timescales , we propose C-RoPE for representing unevenly sampled observation times; we propose EANE for presenting the noisy signal; we propose the multi-view self-distillation architecture for making use of the domain knowledge in light curve science. Our model achieved superior results in 15 out of 16 StarEmbed classification benchmark metrics over hand-crafted features, which is, to the best of our knowledge, the first deep learning model that beats the hand-crafted features in light curve classification tasks. We also show the advantage of our model in few-shot classification over the hand-crafted features. We show the interpretability of our model through the visualization of the StarEmbed feature UMAP, and updates in EANE weights. Our model also shows strong capability in the applications such as similarity search, parameter estimation, anomaly detection, and promising results in non-astronomical domains.  



In this work, we choose the  GLS and phase-folded light curve as the views. These modules are hard to replace with a learned module. This is because the Nyquist frequency goes to infinity for unevenly sampled time series \citep{2018ApJS..236...16V}. This feature makes the significance of periodic signal highly sensitive to the frequency, which makes it hard for differentiable algorithms to learn the periodic features. Meanwhile, the phase-folding operation is also a non-differentiable operation. However, it is possible to find learnable view transformation in other domains. For example, in financial time series analysis, people use fractional differentiation to stabilize financial data, which can be replaced by 1-dimensional convolutional kernels. We would encourage the choice of such view designs to reduce human-induced bias in future works.  

Unfortunately, there is no ``silver bullet'' solution for solving all kinds of time-series processing problems simultaneously \citep{Middlehurst_2024}. This is because time series in different domains are collected and aggregated with the specific domain knowledge and assumptions. For example, light curves represents integrated photon number within the exposure time; financial trade data aggregates the order book within a given time slot; ECG aggregates the heart’s distributed electrical activity through body-surface electrode differences. The time series in different domain contains different underlying dynamics and emergent behavior. The semantic meanings of individual data points and observation patches of observation of the time series are therefore also different. Therefore, although our method achieves several state-of-the-art results on the PYRREGULAR dataset, we suggest that applications to domains beyond light curves should further tailor the architecture by leveraging domain-specific knowledge to design task-specific views and encoders.

\section*{Acknowledgement}
Yicheng Rui is supported by T.D. Lee scholarship. We thank Professor Yuan-Sen Ting and Dr. Kaiming Cui for helpful discussions, and Nabeel Rehemtulla for assistance with the StarEmbed dataset.

\bibliographystyle{plainnat}
\bibliography{example_paper}

\newpage
\section*{Appendix}
\appendix

\section{Generalized Lomb-Scargle periodogram (GLS)\label{sec:GLS}}

The GLS is a widely used method for analyzing the periodic signal in astronomical time series data. It fits the time series $\{t_i,y_i,\sigma_i\}$ using 

\begin{equation}
 \hat{y}(t) = a \sin \omega t + b \cos \omega t + c
\end{equation}
at a given period $P = \frac{2\pi}{\omega}$. By minimizing

\begin{equation}
 \chi^2 (\omega) = \min\sum_i \frac{(y_i-\hat{y}(t))^2}{\sigma_i^2},
\end{equation}
for each $\omega$, the normalized power spectrum is defined as

\begin{equation}
 p(\omega) = \frac{\chi^2_0 - \chi^2 (\omega)}{\chi^2_0},
\end{equation}

where $\chi_0$ is the constant baseline at $\omega = 0$. Its closed-form solution is shown in \citep{Zechmeister09}. In this work, the period $P$ is sampled from 20 minutes to 2000 days evenly in log-space with $10^6$ samples. To save memory, we take the samples with 500 most significant signals and 500 other random samples to feed into the periodogram view.




\section{Model Parameters and Experimental Details}
\label{app:model_parameters_experimental_details}
\providecommand{\modelappendixfigdir}{runs/appendix_figures/mean_pooling_model}

This appendix reports the configuration used for the no-overlap LEAVES pretraining
of the mean-pooling model. The LEAVES manifest excludes duplicated StarEmbed targets and contains 1,056,785 light curves.
The active pretraining views are raw light curve, periodogram, and phase-folded light curve. The group-fusion branch and projection head are disabled.

\begin{table*}[t]
\centering
\caption{Backbone parameterization for the full model.}
\label{tab:appendix_model_parameters}
\begin{tabular}{l l l}
\specialrule{2pt}{1pt}{1pt}
Component & Setting & Value \\
\specialrule{2pt}{1pt}{1pt}
Raw-view encoder & Transformer depth & 4 layers \\
Raw-view encoder & Hidden width & 256 \\
Raw-view encoder & Attention heads & 4 \\
Raw-view encoder & Position encoding & C-RoPE over observation time \\
Raw-view encoder & Numeric embedding & Error-aware Gaussian embedding, 2,048 bins \\
Raw-view encoder & Pooling & Masked mean pooling \\
Raw-view encoder & Output dimension & 128 \\
Raw-view encoder & Parameters & 3,713,664 \\
\hline
Periodogram-view encoder & Transformer depth & 4 layers \\
Periodogram-view encoder & Hidden width & 256 \\
Periodogram-view encoder & Attention heads & 4 \\
Periodogram-view encoder & Position encoding & C-RoPE over $\log_{10} P$ \\
Periodogram-view encoder & Numeric embedding & Log-power embedding, 2,048 bins \\
Periodogram-view encoder & Pooling & Masked mean pooling \\
Periodogram-view encoder & Output dimension & 128 \\
Periodogram-view encoder & Parameters & 3,713,664 \\
\hline
Phase-folded-view encoder & Transformer depth & 4 layers \\
Phase-folded-view encoder & Hidden width & 256 \\
Phase-folded-view encoder & Attention heads & 4 \\
Phase-folded-view encoder & Position encoding & C-RoPE over normalized phase \\
Phase-folded-view encoder & Numeric embedding & Error-aware Gaussian embedding, 2,048 bins \\
Phase-folded-view encoder & Pooling & Masked mean pooling \\
Phase-folded-view encoder & Output dimension & 128 \\
Phase-folded-view encoder & Parameters & 3,713,664 \\
\hline
Backbone & Total parameters & 11,140,992 \\
\specialrule{2pt}{1pt}{1pt}
\end{tabular}
\end{table*}

\begin{table*}[t]
 \centering
 \caption{Pretraining and detached-probe optimization settings for the no-overlap full model.}
 \label{tab:appendix_training_details}
 \scriptsize
\setlength{\tabcolsep}{3pt}
\renewcommand{\arraystretch}{0.95}
 \begin{tabular}{l l}
  \specialrule{2pt}{1pt}{1pt}
  Setting & Value \\
  \specialrule{2pt}{1pt}{1pt}
  Pretraining objective & LeJEPA only, weight 1.0, $\lambda=0.02$; CLIP loss disabled \\
  Regularization & SIGReg with 17 knots, maximum $t=3.0$, projection dimension 256 \\
  Views in loss & Raw, periodogram, phase-folded \\
  Numeric value bins & 2,048 bins for light curve magnitudes and periodogram log-power values \\
  Raw magnitude range & $[-4, 4]$ with error-aware Gaussian numeric embedding \\
  Periodogram log-power range & $[-6, 0]$ \\
  Batch size & 128 per process; 8 DDP processes; effective global batch size 1,024 \\
  Training length & 10,000 optimizer steps; gradient accumulation 1 \\
  Backbone optimizer & AdamW, learning rate $2\times10^{-4}$, weight decay 0.01 \\
  Backbone AdamW parameters & $\beta_1=0.9$, $\beta_2=0.999$, $\epsilon=10^{-8}$ \\
  Backbone gradient clipping & Global norm 1.0 \\
  Mixed precision & Enabled \\
  Data split & Test ratio 0.1, seed 1234 \\
  Logging and checkpoints & Log every 10 steps; save every 10,000 steps \\
  Detached-probe features & Concatenated detached embeddings from the three active views, dimension 384 \\
  Detached-probe optimizer & AdamW, learning rate $10^{-3}$, weight decay 0 \\
  Detached-probe losses & Cross-entropy for 7-class and 10-class labels; labels $<0$ ignored \\
  \specialrule{2pt}{1pt}{1pt}
 \end{tabular}
\end{table*}

\begin{figure*}[t]
 \centering
 \includegraphics[width=0.92\textwidth]{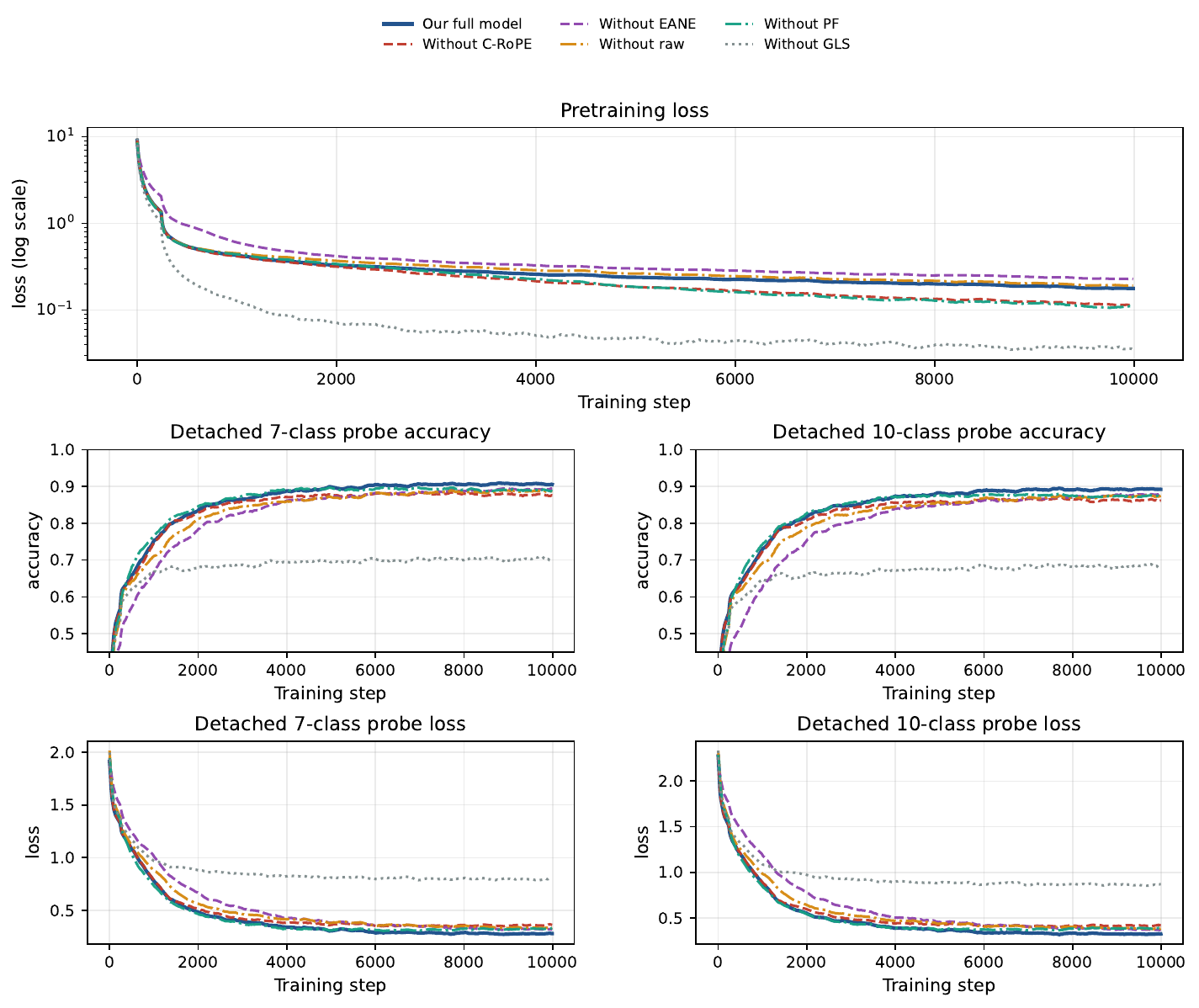}
 \caption{Training curves for the detached embedding probes during pretraining on LEAVES.
 The lines traces show
 a moving average during model training. The probes are optimized on detached concatenated embeddings and do not
 backpropagate into the backbone. }
 \label{fig:appendix_detached_probe_training_curves}
\end{figure*}

\subsection{LEAVES--StarEmbed Deduplication}
\label{app:leaves_starembed_deduplication}

The LEAVES pretraining set and the StarEmbed benchmark are both derived from
ZTF light curves, so target-level deduplication is required before using
StarEmbed as a downstream evaluation benchmark. The filtering is performed at
the astronomical-source level rather than by removing all ZTF targets. In
particular, only LEAVES light curves whose target is crossmatched to a
StarEmbed source are removed; all other ZTF light curves remain in the LEAVES
pretraining corpus.

We first construct the set of StarEmbed evaluation targets from the preprocessed
StarEmbed splits used in this work: train, validation, and test. Each StarEmbed preprocessed file name contains the Pan-STARRS source identifier, and
these identifiers are collected from the split manifests. The resulting
Pan-STARRS source set contains 40,716 unique StarEmbed targets. We then query the StarEmbed crossmatch table, which provides Pan-STARRS source identifier, Gaia source identifier, right ascension, and declination. Restricting this table to the collected StarEmbed sources gives 41,045 catalog rows, 40,589 unique Gaia identifiers, and 40,590 unique coordinate entries with valid coordinates.

LEAVES targets are matched to this StarEmbed source catalog in two passes. The first pass checks whether a LEAVES target name explicitly contains a Gaia identifier of the form \texttt{GaiaID<source\_id>}. If that Gaia identifier is present in the StarEmbed crossmatch table, the LEAVES target is removed. The second pass is a positional crossmatch for LEAVES target names that encode sky coordinates. We parse the target coordinate string as \texttt{JHHMMSS.s+DDMMSS.s} or \texttt{JHHMMSS.s-DDMMSS.s}, convert the right ascension and declination to degrees, and represent both LEAVES and StarEmbed positions as three-dimensional unit vectors on the celestial sphere. A \texttt{cKDTree} is built over the StarEmbed unit vectors, and a LEAVES target is removed if it has any StarEmbed neighbor within 2 arcsec. The chord radius used by the tree query is
\[
r = 2\sin\left(\frac{\theta}{2}\right),
\qquad \theta = 2~\mathrm{arcsec},
\]
with $\theta$ converted to radians. This spherical-vector query avoids right-ascension wrap-around issues and applies the same angular threshold over
the full sky.

The final no-overlap manifest is written by scanning the original LEAVES manifest and retaining only targets that fail both the Gaia-identifier match and the 2-arcsec positional match. In our run, the original LEAVES manifest contains 1,099,827 light curves. The coordinate parser is applicable to 965,231 LEAVES target names; 43,042 light curves are removed by the positional crossmatch, and no additional light curves are removed by the Gaia-identifier pass. The resulting no-overlap LEAVES manifest contains 1,056,785 light curves. This manifest is used for all final LEAVES pretraining and for the re-pretraining of comparison models.

\subsection{Model Training Details}
\label{app:model_training_details}

All LEAVES pretraining experiments in the final comparison use the no-overlap LEAVES manifest, which removes duplicated StarEmbed targets before pretraining. This prevents the frozen StarEmbed evaluation split from appearing in the
self-supervised pretraining corpus. Unless otherwise stated, models are trained for 10,000 optimizer steps with seed 1234, batch size 128 per process, eight DDP processes, gradient accumulation 1, gradient clipping at global norm 1.0, and AdamW with learning rate $2\times10^{-4}$, weight decay 0.01, betas
$(0.9,0.999)$, and $\epsilon=10^{-8}$. The detached probes use concatenated
frozen embeddings, learning rate $10^{-3}$, and no weight decay.

\begin{itemize}
 \item \textbf{Our model}: Our main model uses three numeric views: the raw
 light curve, the Lomb--Scargle periodogram, and the phase-folded light
 curve. The group branch and projection head are disabled. Each view encoder has width 256, four transformer blocks, four attention heads, and a 128-dimensional output embedding. The raw and phase-folded views use time/phase C-RoPE and error-aware numeric embeddings; the periodogram view uses C-RoPE over periodogram coordinates. The final sequence
 representation is obtained by masked mean pooling over valid transformer
 tokens. The pretraining objective is LeJEPA with weight 1.0 and
 $\lambda=0.02$, together with SIGReg using 17 knots, maximum $t=3.0$, and
 projection dimension 256. CLIP loss is disabled for the main model. The
 three 128-dimensional view embeddings are concatenated for detached probing and StarEmbed extraction, giving a 384-dimensional representation. The backbone has 11,140,992 trainable parameters.

 \item \textbf{Astromer-2}: Astromer-2 is re-pretrained on the same no-overlap LEAVES manifest rather than using the original public checkpoint. To make the comparison size-matched, its encoder is resized to match the encoder scale used by our model: $d_{\mathrm{model}}=256$, four transformer layers, four attention heads, feed-forward dimension 1024, dropout 0.1, and RoPE base 10000. The Astromer-2 masking configuration
 uses probed fraction 0.5, masked fraction 0.3, and random fraction 0.1. The input sequence length is capped at 1000 for LEAVES pretraining, with  period search range $[0.006944444,2000]$ over 1000 grid points. The model
 uses the shared 10,000-step AdamW training recipe above, with batch size 128, train fraction 0.9, point-drop augmentation probability 0.1, and maximum dropped-point fraction 0.10. StarEmbed extraction uses sequence length 200, z-score normalization, batch size 256, and the final checkpoint.

 \item \textbf{FALCO}: FALCO is also re-pretrained from scratch on the same no-overlap LEAVES manifest and is resized to the encoder scale used by our model: $d_{\mathrm{model}}=256$, four transformer layers, four attention heads, feed-forward dimension 1024, dropout 0.1, and RoPE base 10000. Error bars are included in the input representation, the Huber loss uses $\delta=1.0$, the maximum accepted error bar is 5.0, and times are normalized to start at zero. It uses the same 10,000-step AdamW recipe,
 batch size 128, train fraction 0.9, mixed precision, gradient clipping at 1.0, and point-drop augmentation probability 0.1 with maximum dropped-point fraction 0.10. StarEmbed features are extracted from the final checkpoint with sequence length 200, batch size 256, and mean pooling.

 \item \textbf{Ablation models}: The ablation suite is trained with the same no-overlap data, masked mean pooling, encoder size, optimizer, and training length as the main model unless the ablated component changes the active views or objective. We train one-component ablations that remove C-RoPE, remove the error-aware numeric embedding, remove the raw-view branch, remove the phase-folded-view branch, remove the GLS periodogram branch, or replace the LeJEPA objective with the symmetric CLIP objective between raw light-curve view and other views. These ablations
 isolate whether the gains come from temporal rotary encoding, uncertainty
 modeling, individual time-series views, or the multi-view predictive loss.
\end{itemize}

\subsection{Experiment setup}
The detailed setup of StarEmbed and PYRREGULAR experiments are as follows
\begin{itemize}
\item \textbf{StarEmbed}: We follow the evaluation protocol of the StarEmbed benchmark and evaluate frozen
representations on the official train, validation, test, and anomaly splits. For our models, the $g$- and $r$-band
light curves are encoded separately and the two band-level embeddings are concatenated. 

For the MLP probe, embeddings are standardized with a \texttt{StandardScaler} fitted on the training split only.
The probe is a three-layer MLP with hidden dimensions $[1024,512,256]$, ReLU activations, and inverse-frequency
class weighting. We sweep batch size in $\{32,64,128,256\}$, learning rate in $\{10^{-2},10^{-3},10^{-4}\}$, and
dropout in $\{0.0,0.1\}$. Each run is trained for at most 50 epochs with early stopping on validation loss using
patience 3. The checkpoint with the best validation loss is used for test evaluation.

For $k$-NN, the training and validation splits are concatenated to form the training set like the original StarEmbed benchmark. Features are
standardized using statistics from this combined set, and a $k$-nearest-neighbor classifier with $k=5$ is fitted.

For logistic regression, we use the same train-plus-validation protocol and feature standardization, with \texttt{LogisticRegression} using \texttt{max\_iter=5000}, balanced class weights, and
parallel fitting. 

For Random Forest, we tune hyperparameters on the original train/validation split using macro-F1 as the selection metric. The grid contains \texttt{n\_estimators} $\in \{100,200,500\}$, \texttt{max\_depth} $\in \{\texttt{None},10,20,30\}$, and \texttt{min\_samples\_split} $\in \{2,5,10\}$. After selecting the best configuration, the random forest model is retrained on train plus validation and evaluated on the test set. Random Forest results are averaged over seeds $42$, $100$, and $200$. We report accuracy, macro recall, macro precision, and macro F1 in percentile.

\item \textbf{PYRREGULAR}: We evaluate transfer to irregular multivariate time-series classification using the PYRREGULAR benchmark. The default suite contains \texttt{Mimic3} (MI3), \texttt{Ldfpa} (LPA), \texttt{Pamap2} (PAM),
   \texttt{Animals} (AN), \texttt{GeolifeSupervised} (GS), \texttt{Seabirds} (SE), \texttt{Garment} (PGE), \texttt{ABF},
   \texttt{Vehicles} (VE), \texttt{Physionet2012} (P12), \texttt{Physionet2019} (P19), and \texttt{Taxi} (TA). Each dataset is loaded
   through the PYRREGULAR irregular-series interface and converted to dense arrays with the provided train/test
   split. Non-test samples are used for training downstream classifiers, and samples marked as test are used only for
   final evaluation.

   For representation extraction, each channel of a multivariate series is treated as one univariate irregular light
   curve. A channel input is represented as $(t,x,\sigma,m)$, where $t$ is time, $x$ is the observed value, $\sigma$
   is a constant error value set to $0.1$, and $m$ is the validity mask. In the current fine-tuning setting, values are
   not Chronos-normalized, time is represented with the relative-time strategy, and each channel is capped at 1000
   valid points. The raw-view encoder is used for embedding extraction; periodogram, phase-folded, and group views
   are disabled during inference. Channel embeddings from the same sample are aggregated by mean pooling, producing
   one fixed-dimensional feature vector per multivariate example.

   We also perform dataset-specific adaptation from the checkpoint of our model. For each PYRREGULAR dataset, the model is
   initialized from our pretrained weights from LEAVES, and adapted for 10 epochs on the non-test split using the same self-supervised multi-view objective as fine-tuning. The adaptation uses full-model fine-tuning, DataParallel over up
   to 8 GPUs, automatic global batch-size selection from
   \begin{equation*}
  \begin{aligned}
   \{4096,3072,2048,1536,1024,768,512,384,256,192,128,64,32,16,8\},
  \end{aligned}
   \end{equation*}
and AdamW with the optimizer parameters inherited from the LEAVES pretraining configuration. Dataset-specific numeric ranges, RoPE periods, periodogram value ranges, and inference period bounds are estimated from the training split before adaptation.
Downstream PYRREGULAR classifiers are trained on standardized extracted embeddings. Logistic regression uses \texttt{max\_iter=5000}, balanced class weights, and \texttt{random\_state=42}. The $k$-NN classifier uses $k=\min(5,N_{\mathrm{train}})$, and the Random Forest uses 500 trees with balanced subsample class weights. We report dataset-level accuracy, macro precision, macro recall, and macro F1, and aggregate performance is computed as the arithmetic mean across datasets.
  \end{itemize}

\section{Layer-wise visualization for StarEmbed test set embeddings}
\begin{figure*}

    \includegraphics[width = 1 \linewidth]{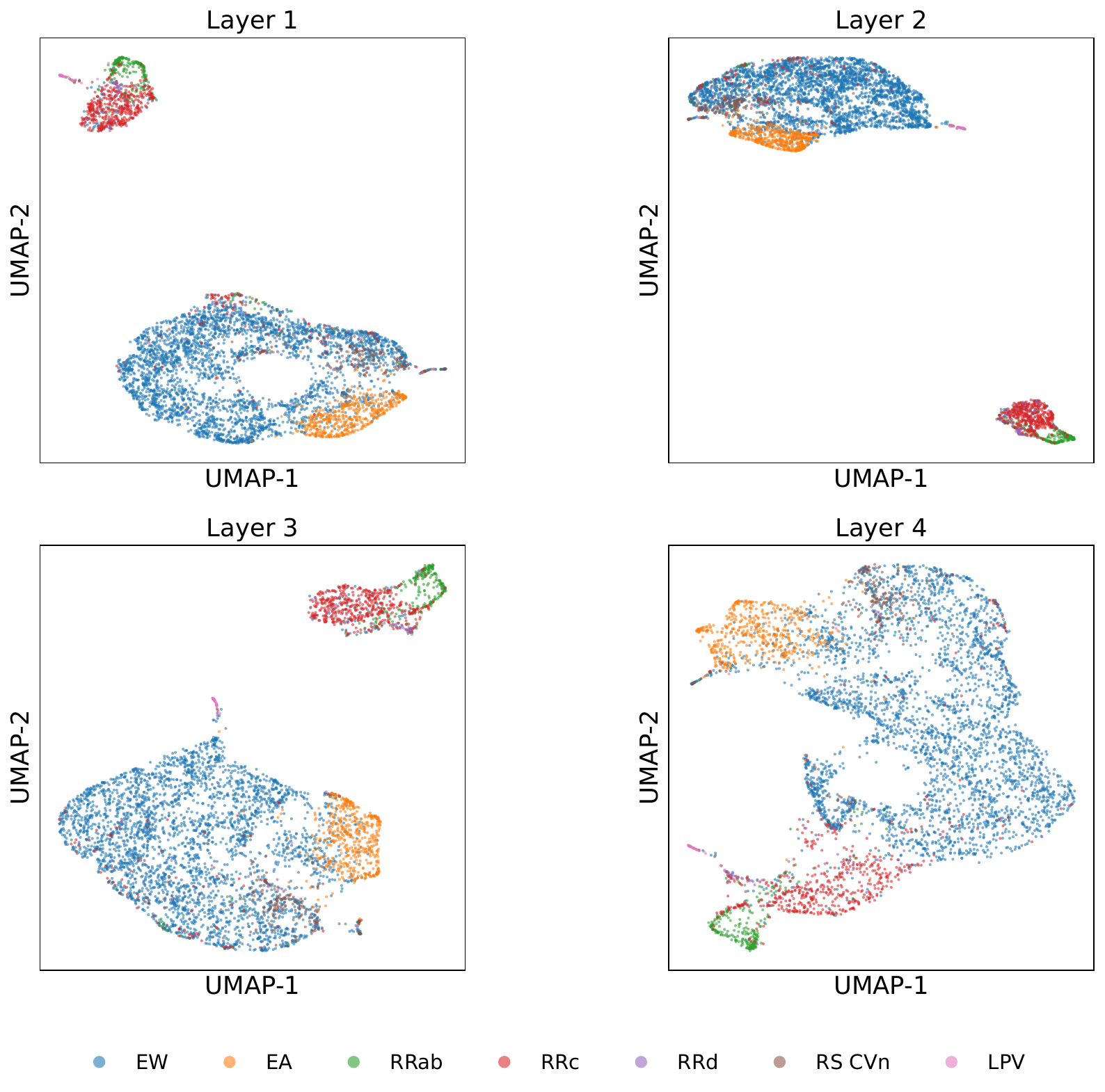}
        \caption{Layer-wise UMAP visualization  of StarEmbed test set from mean-pooling raw light-curve view encoder \label{fig:rawumap}}
\end{figure*}

\begin{figure*}

    \includegraphics[width = 1 \linewidth]{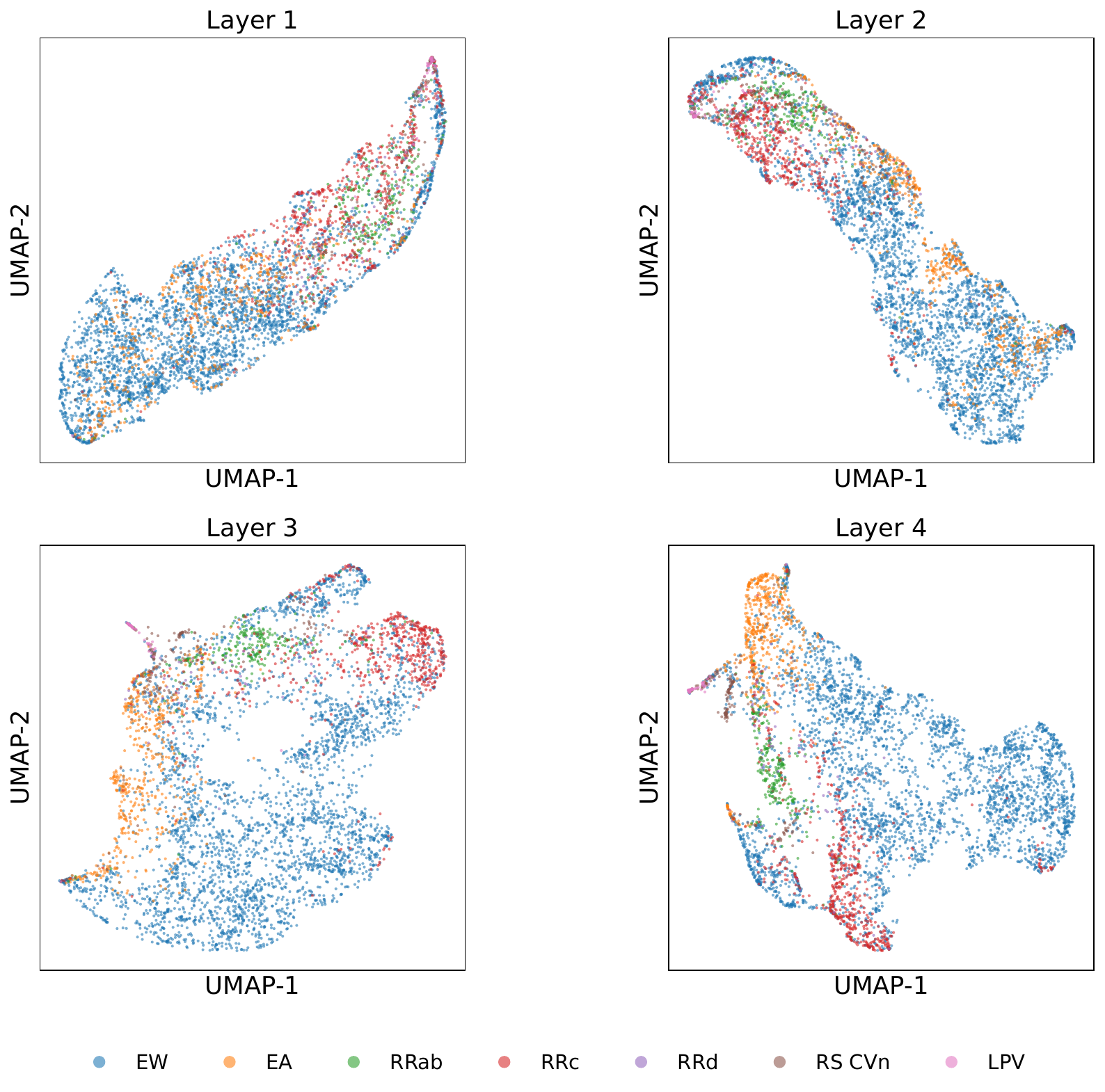}
        \caption{Layer-wise UMAP visualization  of StarEmbed test set from the mean-pooling of GLS view encoder \label{fig:glsumap}}
\end{figure*}

\begin{figure*}

    \includegraphics[width = 1 \linewidth]{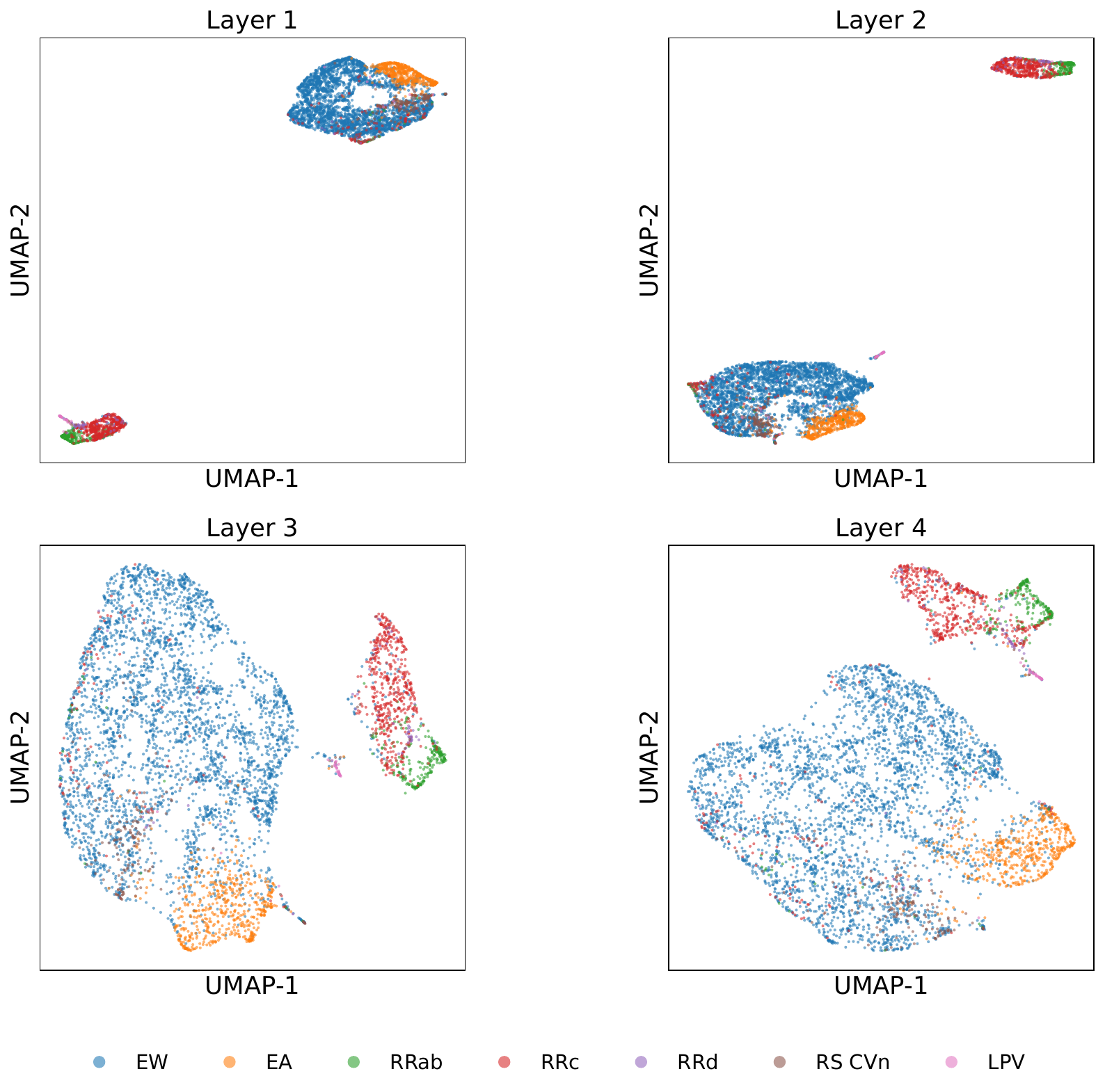}
        \caption{Layer-wise UMAP visualization  of StarEmbed test set from the mean-pooling of phase-folded view encoder \label{fig:pfumap}}
\end{figure*}

For the StarEmbed test set, the mean aggregation over time ticks for different view and different layer are visualized using a UMAP projection in Fig. \ref{fig:rawumap}, Fig. \ref{fig:glsumap}, and Fig. \ref{fig:pfumap}.  

\section{Distribution of L2 distance between different classes\label{app:l2}}

\begin{figure*}
\centering
\includegraphics[width=\linewidth]{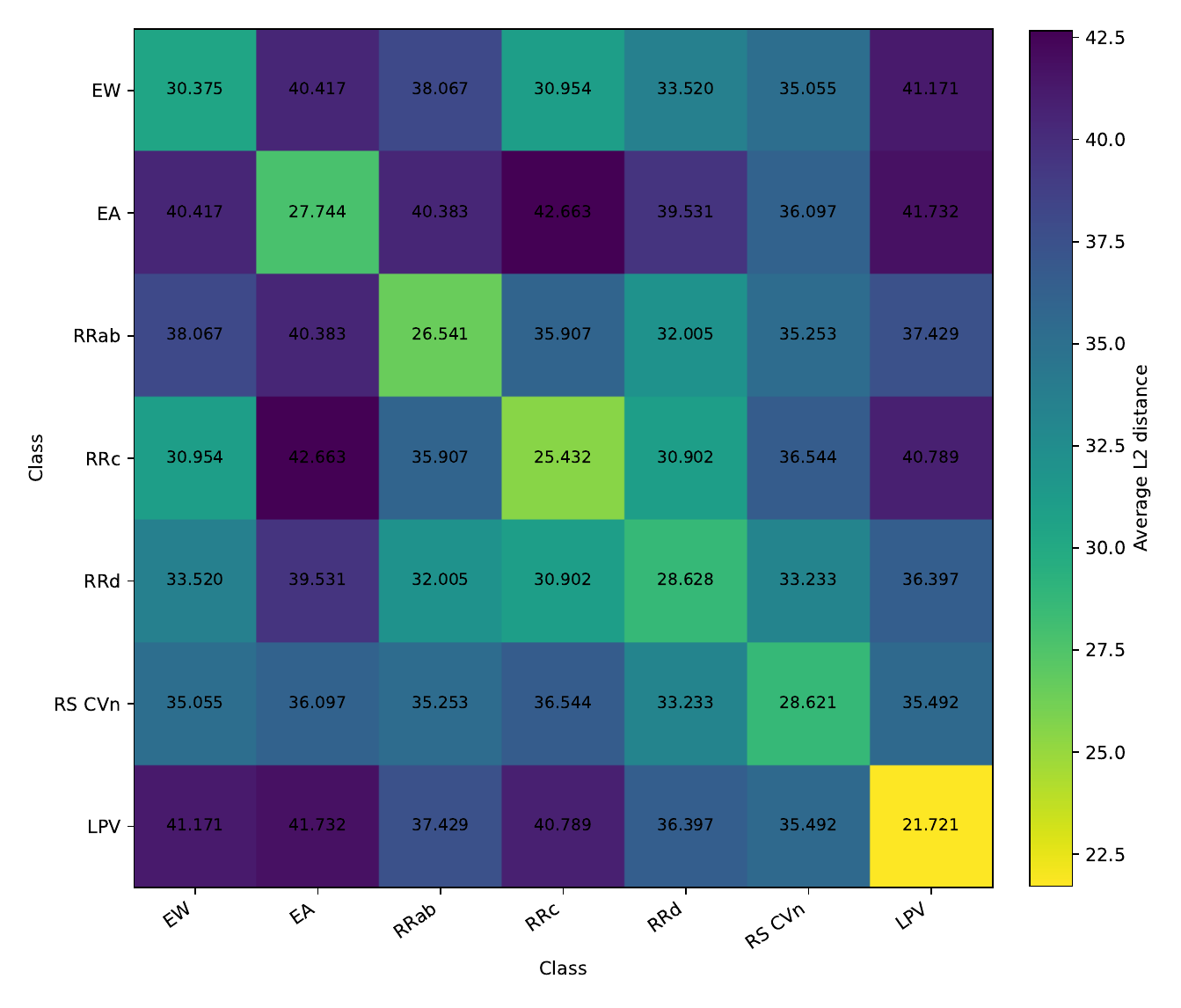}
\caption{Class-wise mean L2 distance of embeddings for StarEmbed test set \label{fig:l2distavg}}
\end{figure*}

\begin{figure*}
\centering
\includegraphics[width=\linewidth]{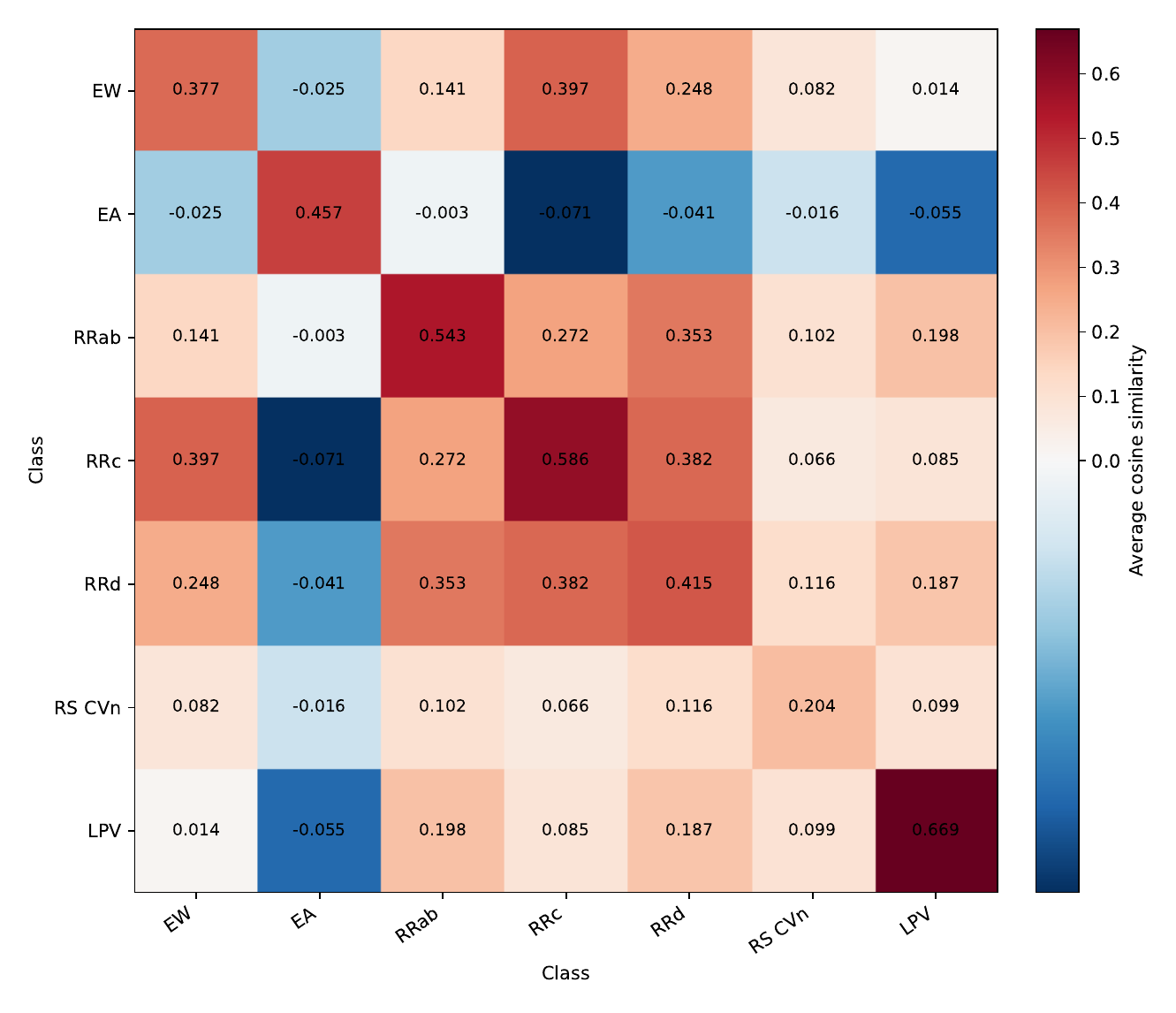}
\caption{Class-wise mean cosine similarity of embeddings for StarEmbed test set \label{fig:cosdistavg}}
\end{figure*}

We visualized the average class-wise L2 distance of embedding for StarEmbed test set in Fig. \ref{fig:l2distavg}. It is shown that the L2 distance between the samples in the same class is always shorter than the distances among different classes, which shows the good clustering property of our model. Meanwhile, the inter-class cosine similarities are shown in Fig. \ref{fig:cosdistavg}. The average cosine similarity between RRc- and RRd-type variable stars is higher than the intra-class similarity among RS CVn-type variable stars. This suggests that, for our model, L2 distance is more appropriate than cosine similarity for similarity search.

\section{Domain-specific adaptation for PYRREGULAR dataset}
\label{app:pyrregularadaption}

To further test whether the proposed framework can be adapted beyond
astronomical light curves, we evaluate a general irregular-time-series variant
on PYRREGULAR. PYRREGULAR differs from StarEmbed and LEAVES in both data semantics and observation structure: each sample can contain multiple sensor or clinical channels, the time axis is not necessarily astronomical observing time, and the target labels describe heterogeneous domains such as clinical state, human activity, animal movement, or trajectory type. We therefore use a domain-adapted configuration instead of assuming that the astronomy-optimized model is universally optimal. Unlike our model used for the main StarEmbed benchmark, this variant uses attention-statistics pooling and keeps the symmetric CLIP term between raw--periodogram and raw--phase-folded embeddings. 

The attention-statistics pooling is as follows. 
 Each numeric view is first converted into a sequence of token embeddings and processed by a
  continuous-position transformer. For a given view, let the final transformer output after the
  last block and final layer normalization be
  \[
      X \in \mathbb{R}^{B \times S \times D},
  \]
  where \(B\) is the batch size, \(S\) is the sequence length of that view, and \(D\) is the
  transformer width. In our EXP10 configuration, \(D=256\). Attention-statistics pooling maps
  this variable-length token sequence to a fixed-dimensional view embedding.

  The pooling layer also receives a validity mask
  \[
      M \in \{0,1\}^{B \times S},
  \]
  where \(M_{b,s}=1\) indicates that token \(s\) in sample \(b\) is valid, and \(M_{b,s}=0\)
  indicates padding or an invalid observation. For raw and phase-folded light curves, this mask
  comes directly from the fourth input channel of the precomputed light-curve tensor:
  \[
      \mathrm{lc}_{b,s,:} = [t_{b,s}, m_{b,s}, \sigma_{b,s}, M_{b,s}],
  \]
  and similarly for the phase-folded view. For the periodogram view, all periodogram tokens are
  treated as valid, so the mask is an all-ones tensor with shape \(B \times S\).

  Given \(X\) and \(M\), the pooling layer computes four masked summaries: a learned
  attention-weighted feature, a masked mean, a masked standard deviation, and a masked maximum.

  First, each token is assigned a scalar attention score using a two-layer scoring network:
  \[
      a_{b,s} = w_2^\top \tanh(W_1 X_{b,s} + b_1) + b_2,
  \]
  where \(a \in \mathbb{R}^{B \times S}\). Invalid tokens are suppressed before softmax:
  \[
      \tilde{a}_{b,s} =
      \begin{cases}
      a_{b,s}, & M_{b,s}=1,\\
      -\infty, & M_{b,s}=0.
      \end{cases}
  \]
  The attention weights are then
  \[
      \alpha_{b,s} =
      \frac{\exp(\tilde{a}_{b,s})}{\sum_{j=1}^{S}\exp(\tilde{a}_{b,j})}.
  \]
  After softmax, invalid positions are explicitly zeroed and the weights are renormalized for
  numerical stability:
  \[
      \alpha_{b,s} \leftarrow
      \frac{\alpha_{b,s} M_{b,s}}{\sum_{j=1}^{S} \alpha_{b,j} M_{b,j} + \epsilon}.
  \]
  The learned attention-pooled feature is
  \[
      h^{\mathrm{attn}}_b = \sum_{s=1}^{S} \alpha_{b,s} X_{b,s}
      \in \mathbb{R}^{D}.
  \]

  The masked mean is computed as
  \[
      h^{\mathrm{mean}}_b =
      \frac{\sum_{s=1}^{S} M_{b,s} X_{b,s}}
           {\max\left(\sum_{s=1}^{S} M_{b,s}, 1\right)}
      \in \mathbb{R}^{D}.
  \]

  The masked standard deviation is computed per feature dimension:
  \[
      h^{\mathrm{std}}_b =
      \sqrt{
      \frac{\sum_{s=1}^{S} M_{b,s}
      \left(X_{b,s} - h^{\mathrm{mean}}_b\right)^2}
      {\max\left(\sum_{s=1}^{S} M_{b,s}, 1\right)}
      + \epsilon}
      \in \mathbb{R}^{D}.
  \]

  The masked maximum is computed by replacing invalid token features with a very
  negative value before taking the maximum:
  \[
      h^{\mathrm{max}}_{b,d} =
      \max_{s: M_{b,s}=1} X_{b,s,d},
      \qquad
      h^{\mathrm{max}}_b \in \mathbb{R}^{D}.
  \]

  The four summaries are concatenated:
  \[
      h_b =
      \left[
      h^{\mathrm{attn}}_b ;
      h^{\mathrm{mean}}_b ;
      h^{\mathrm{std}}_b ;
      h^{\mathrm{max}}_b
      \right]
      \in \mathbb{R}^{4D}.
  \]
  In our model parameter setting, this gives \(h_b \in \mathbb{R}^{1024}\), since \(D=256\).

  Finally, this concatenated vector is passed through a fusion MLP:
  \[
      z_b =
      W_o \, \mathrm{Dropout}
      \left(
      \mathrm{GELU}(W_h h_b + b_h)
      \right)
      + b_o,
  \]
  where the hidden width is 256, the output dimension is 128.

  Meanwhile, the  combined loss used by the adaptation model is written as
  \begin{equation}
\begin{aligned}
    \mathcal{L}_{\rm total} &= \alpha_{\rm LeJEPA}\mathcal{L}_{\rm LeJEPA} + \alpha_{\rm CLIP} \mathcal{L}_{\rm CLIP} \\
    \mathcal{L}_{\rm LeJEPA} &= \lambda \mathcal{L}_{\rm SIGReg} + (1 - \lambda) \mathcal{L}_{\rm inv} \\
    \mathcal{L}_{\rm inv} &= \frac{1}{V} \sum_{v=1}^{V} \| z_v - \bar{z} \|^2 \\
    \mathcal{L}_{\rm SIGReg} &= \mathbb{E}_{\bar{z}} \left[ \int \left| \hat{\phi}_{\rm data}(t; \bar{z}) - \phi_{\rm target}(t) \right|^2 w(t) \, dt \right] \\
    \mathcal{L}_{\rm CLIP} &= \frac{1}{|\text{pairs}|} \sum_{(a,b) \in \text{pairs}} \mathcal{L}_{\rm pair}(z_a, z_b) \\
    \mathcal{L}_{\rm pair}(z_a, z_b) &= \frac{1}{2} \left( \mathcal{L}_{CE}\left(\frac{z_a z_b^\top}{\tau}, y\right) + \mathcal{L}_{CE}\left(\frac{z_b z_a^\top}{\tau}, y\right) \right),
\end{aligned}
\end{equation}
 where temperature $\tau=0.2$, $\alpha_{\rm LeJEPA} = \alpha_{\rm CLIP} = 1$.

  The comparison between our model and its PYRREGULAR-adapted variant is shown in Table \ref{tab:PYRREGULAR_domain_adaptation}. It is interesting to see that the variant model beats previous SOTA results on 5 out of 12 PYRREGULAR datasets, while it still has worse performance on LPA and MI3 datasets than our model. The variant model achieves the improvement on some datasets with worse performance on others, which indicates the domain-specific requirements for time series.

\begin{table*}[t]
    \caption{PYRREGULAR macro-F1 comparison between the previous SOTA, our model, and its PYRREGULAR-adapted variant.}
    \label{tab:PYRREGULAR_domain_adaptation}
    \centering
    \begin{tabular}{l|lc|lc|lc}
    \hline
    Dataset
    & Prev. SOTA & Prev. F1
    & No-overlap head & No-overlap F1
    & Adapted head & Adapted F1 \\
    \hline
    ABF & NCDE & 0.41 & Linear & 0.37 & Linear & 0.55 \\
    AN & ROCKET & 0.90 & $k$-NN & 0.58 & $k$-NN & 0.52 \\
    PGE & BRITS & 0.78 & Linear & 0.78 & $k$-NN & 0.78 \\
    GS & NCDE & 0.52 & Linear & 0.20 & Linear & 0.20 \\
    LPA & BORF & 0.73 & Linear & 0.82 & Linear & 0.77 \\
    MI3 & RIFC & 0.56 & Linear & 0.69 & Linear & 0.61 \\
    PAM & ROCKET & 0.66 & Linear & 0.49 & Linear & 0.74 \\
    P12 & RIFC & 0.63 & Linear & 0.58 & Linear & 0.61 \\
    P19 & LGBM & 0.75 & $k$-NN & 0.50 & $k$-NN & 0.57 \\
    SE & RIFC & 0.82 & Linear & 0.67 & Linear & 0.67 \\
    TA & LGBM & 0.77 & Linear & 0.35 & Linear & 0.36 \\
    VE & BORF & 0.87 & Linear & 0.84 & $k$-NN & 0.64 \\
    \hline
    \end{tabular}
    \end{table*}
    
\section{Algorithm of Domain-Informed Multi-view Self-Distillation for time series\label{app:alg}}
Algorithm~\ref{alg:domain_informed_multiview} summarizes the proposed framework. 
The algorithm is domain-general: it only assumes that domain knowledge can provide 
multiple semantics-preserving views of the same irregular time series. In this work, 
we instantiate these views as the raw light curve, the GLS periodogram, and the 
phase-folded light curve.

\begin{algorithm*}[t]
\caption{Domain-Informed Multi-View Self-Distillation}
\label{alg:domain_informed_multiview}
\begin{algorithmic}[1]
\Require Irregular time series batch $\mathcal{B}=\{(t_i,x_i,\sigma_i)\}_{i=1}^{N}$;
domain-informed view functions $\{g_v\}_{v=1}^{V}$;
view encoders $\{f_{\theta_v}\}_{v=1}^{V}$;
projection or pooling heads $\{h_{\psi_v}\}_{v=1}^{V}$;
LeJEPA weight $\lambda$.
\Ensure Self-supervised representation model $\{f_{\theta_v},h_{\psi_v}\}_{v=1}^{V}$.

\For{each training step}
    \State Sample a mini-batch $\mathcal{B}$ of irregular time series.
    \For{each view $v=1,\ldots,V$}
        \State Construct a domain-informed view:
        \[
            X^{(v)} \gets g_v(\mathcal{B}).
        \]
        \Comment{For light curves: raw, GLS periodogram, phase-folded curve.}

        \State Encode time/value/uncertainty tokens:
        \[
            H^{(v)} \gets f_{\theta_v}(X^{(v)}).
        \]
        \Comment{Use C-RoPE for irregular coordinates and EANE when uncertainties are available.}

        \State Pool token features into a view-level embedding:
        \[
            z^{(v)} \gets h_{\psi_v}(H^{(v)}).
        \]
    \EndFor

    \State Compute the multi-view centroid:
    \[
        \bar{z} \gets \frac{1}{V}\sum_{v=1}^{V} z^{(v)}.
    \]

    \State Compute the view-alignment loss:
    \[
        \mathcal{L}_{\mathrm{inv}}
        \gets
        \frac{1}{V}\sum_{v=1}^{V}
        \left\|z^{(v)}-\bar{z}\right\|_2^2.
    \]

    \State Compute the SIGReg regularization loss:
    \[
        \mathcal{L}_{\mathrm{SIGReg}}
        \gets
        \mathbb{E}_{\bar{z}}
        \left[
        \int
        \left|
        \hat{\phi}_{\mathrm{data}}(s;\bar{z})
        -
        \phi_{\mathrm{target}}(s)
        \right|^2
        w(s)\,ds
        \right].
    \]

    \State Compute the total LeJEPA loss:
    \[
        \mathcal{L}_{\mathrm{LeJEPA}}
        \gets
        (1-\lambda)\mathcal{L}_{\mathrm{inv}}
        +
        \lambda \mathcal{L}_{\mathrm{SIGReg}}.
    \]

    \State Update all encoder and pooling parameters by minimizing
    $\mathcal{L}_{\mathrm{LeJEPA}}$.
\EndFor

\State \Return Trained encoders and view embeddings
$\{z^{(v)}\}_{v=1}^{V}$ for downstream tasks.
\end{algorithmic}
\end{algorithm*}

\end{document}